\documentclass[english,reprint,longbibliography]{revtex4-1}
\usepackage[T1]{fontenc}
\usepackage[latin9]{inputenc}
\usepackage{amsmath}
\usepackage{graphicx}
\usepackage{esint}

\makeatletter
\usepackage{enumitem}

\makeatother

\usepackage{babel}
\begin{document}

\title{Maxwell's demon and the management of ignorance in stochastic thermodynamics}

\author{Ian J. Ford}

\address{Department of Physics and Astronomy and London Centre for Nanotechnology,
University College London, Gower Street, London WC1E 6BT, U.K.}
\begin{abstract}
It is nearly 150 years since Maxwell challenged the validity of the
second law of thermodynamics by imagining a tiny creature who could
sort the molecules of a gas in such a way that would decrease entropy
without exerting any work. The \emph{demon} has been discussed largely
using thought experiments, but it has recently become possible to
exert control over nanoscale systems, just as Maxwell imagined, and
the status of the second law has become a more practical matter, raising
the issue of how measurements manage our ignorance in a way that can
be exploited. The framework of stochastic thermodynamics extends macroscopic
concepts such as heat, work, entropy and irreversibility to small
systems and allows us explore the matter. Some arguments against a
successful demon imply a second law that can be suspended indefinitely
until we dissipate energy in order to remove the records of his operations.
In contrast, under stochastic thermodynamics the demon fails because
on average more work is performed upfront in making a measurement
than can be extracted by exploiting the outcome. This requires us
to exclude systems and a demon that evolve under what might be termed
self-sorting dynamics, and we reflect on the constraints on control
that this implies while still working within a thermodynamic framework.
\end{abstract}
\maketitle

\section{Introduction\label{sec:Introduction}}

In recent years considerable progress has been made in extending the
familiar ideas of thermodynamics to systems at the smallest scales.
According to the framework of `stochastic thermodynamics', the key
steps are to recognise that the complicated evolution of a small system
coupled in some way to its environment may be represented (if only
approximately) as a random walk, and that the probabilistic character
of the walk can be used to define a dynamical quantity, the stochastic
entropy production, that provides a bridge between mechanics and thermodynamics
\cite{Seifert12,sekimoto2}. With such ingredients it has been possible
to cast the second law of thermodynamics and its implications as `\emph{more
what you'd call \textquotedbl{}guidelines\textquotedbl{} than actual
rules}' \cite{Barbossa}; a set of statistical tendencies in line
with the point of view ultimately offered by Boltzmann \cite{Cercignani98}.

These developments promise to offer practical implications as we learn
how to recognise thermodynamic features in the behaviour of small
systems amid the prevailing fluctuations. Furthermore, the framework
has brought clarity to a number of issues in statistical physics,
one of the oldest being that of Maxwell's demon \cite{Leff03}. The
purpose of this article is to summarise the present position on the
demon, and to reflect on the connection between the perception of
system detail, uncertainty in dynamical evolution, and the production
of entropy \cite{fordbook,Parrondo15}.

The second law famously declares that the total entropy of the world
cannot decrease, a statement that is equally famously incompatible
with the supposed time reversal invariance of the underlying dynamics
of its microscopic components. This problem was apparent to James
Clerk Maxwell in the 1860s when he presented the demon in a thought
experiment that, he felt, demonstrated that the second law could potentially
be broken with the right sort of manipulation \cite{Maxwell71}. Even
without such efforts, he considered the law could only be true `on
average' in the course of the natural evolution of the world.

The demon has continued to be discussed up to the present day. This
is possibly because the concept of entropy, and the status of the
second law, have often seemed hard to pin down, as indeed have the
rules that are imagined to apply to demonic activity. Demons, or more
prosaically, devices that can measure properties of a system and then
exploit the findings, have been conceived with different capabilities
and working for and against a variety of interpretations of the second
law. A consensus has been hard to establish.

One of the strengths of stochastic thermodynamics, however, is that
it seems to offer an intuitive and appealing resolution to some of
the issues. To start with, an observer's ignorance of some of the
details of the world is accepted as a key factor in the underlying
dynamics, an inevitable consequence of having to make do with incomplete
prior measurements, and the evolution of this ignorance is identified
with the statistical expectation or mean of the stochastic entropy
production. Furthermore, since this entropy production can be linked
to the dynamics, it becomes clear how its development can be managed
through suitable mechanical actions, either externally imposed or
internally programmed or autonomous, whereby ignorance of certain
features about the evolution might be reduced. This can potentially
be associated with the conversion of environmental heat into mechanical
work, transferred by the system to a potential energy store, but implications
are attached. In short, we can discuss the idea of the control of
a system through feedback, and place it within a context of thermodynamics
and the second law.

We shall reflect on ignorance in Section \ref{sec:Ignorance}, and
discuss the role of Maxwell's demon in thought experiments in Section
\ref{sec:The-purpose-of}, together with recent practical demonstrations
of systems that appear to carry out the actions of a demon. In Section
\ref{sec:Stochastic-thermodynamics-and} we introduce the basic ideas
of stochastic thermodynamics and the form of second law it offers,
and in Section \ref{sec:The-management-of} we discuss conceptually
how we can link the exploitation of measurements with the management
of ignorance. A demonstration of how this can be achieved using a
simple system with feedback informed by measurements is given in Section
\ref{sec:A-simple-model}, and with further detail in Appendix \ref{sec:mathematical-detail}.
We reflect on the rules we implicitly impose upon demons in Section
\ref{sec:Rules-for-demons} and note that systems could possess self-sorting
dynamics that allow them to reduce their entropy, though it is unclear
whether this would correspond to a thermodynamic context.  Finally,
in Section \ref{sec:Conclusions} we summarise and contrast the various
positions that can be taken on the issues of measurement and exploitation,
and suggest that stochastic thermodynamics provides the clearest yet
presentation of the capabilities and limitations of Maxwell's demon,
and the status and meaning of the second law.

\section{Ignorance\label{sec:Ignorance}}

It is a sad condition of life that we have little idea about the state
of the world around us or a clear view of what the future might hold,
though we manage to cope most of the time. In science, as in life,
our perception is incomplete because of a limited ability to make
measurements; our predictions of the future rely on this perception
as well as on previous experience, and they are consequently hazy.
Statistical physics, and its precursor thermodynamics, were developed
to provide guidance in spite of such conditions of ignorance and they
have been remarkably, and perhaps surprisingly, successful.

We have, of course, devised tools to reveal details of systems on
scales that would otherwise be hard to perceive, and we have used
what we have found to create better models of how a given situation
will evolve. Nevertheless, it is necessary to accept that personal
uncertainty remains a feature of our dealings with the world, and
we should reflect briefly on what this implies.

If we confine ourselves to classical, rather than quantum phenomena,
the world is in principle a rather straightforward place. If only
we knew the equations of motion and the coordinates (the microstate)
of all the component particles and fields, it seems that we could
fully predict the future and retrodict the past, as imagined by Laplace
two centuries ago \cite{Laplace}. We do not know all this, of course,
and neither could we carry out the computations. In actual fact we
are obliged to consider relatively small parts of the world at a time.
We naturally select those parts that conform to a simple profile:
that they should be predictable without obliging us to specify all
the details. This is what is meant by thermodynamic modelling.

Some dynamical systems will evolve in a way that is rather independent
of neglected features such as initial conditions. For example, we
might imagine a subset of the world to have dynamics possessing what
are called \emph{fixed points} or \emph{attractors} \cite{Strogatz-book}.
Asymptotically, such a subsystem will tend towards a certain behaviour
that could be static, cyclic or even chaotic with particular emergent
features. These outcomes will arise with little or no dependence on
the initial conditions of the system and the rest of the world: an
ignorance of the initial microstate does not prevent the emergence
of a greater clarity in the future. But these are not the most common
types of system we encounter in the real world. It is more natural
that initial uncertainty regarding the microstate is amplified as
time moves forward, though it possibly might eventually reach some
sort of ceiling \cite{Ford15aa}. In these cases, the poorly specified
interactions between the system and its environment are typically
of a kind that make the system evolve with great sensitivity to the
neglected details of the initial conditions.

In such a situation, it is clear that we might have to use a probability
distribution over system microstates to represent our uncertainty
and to provide an assessment of future behaviour. Furthermore, even
if we are certain about the system microstate at the present time,
we would tend to become more ignorant as time advances. Now, there
is a mathematical property of a probability distribution can provide
a measure of uncertainty or ignorance, and it is called the Shannon
entropy. If the probabilities over the discrete microstates of the
system are $p_{i}$, the Shannon entropy is defined as $S_{I}=-\sum_{i}p_{i}\ln p_{i}$.
This quantity has the features we intuitively require of a measure
of ignorance, for example it equals zero if the microstate is precisely
known (one of the $p_{i}$ is unity with all others zero), and if
the system consists of independent parts (such that the $p_{i}$ factorise
into probabilities for the microstates of each subsystem), the Shannon
entropy then reduces to a sum of appropriately defined contributions
from each part. The Shannon entropy takes a maximum value if we are
entirely uncertain as to the system microstate, which we would represent
by the assignment of equal probabilities to all of them \cite{Jaynes03,Grandy-book2012}.
For a continuum of system microstates, described by a probability
density function (pdf) rather than a discrete distribution, we extend
the definition to write
\begin{equation}
S_{I}=-\int dx\ p(x)\ln p(x).\label{eq:1-1}
\end{equation}
An obvious flaw here is that the argument of the logarithm is not
dimensionless, but this form is adequate for computing the shift in
Shannon entropy when the pdf describing a system changes.

At this point it is worth commenting that the word `ignorance' is
not commonly used in connection with the Shannon entropy. The more
usual technical term is `information', employed in the sense that
it is a measure of what the observer does \emph{not} know about the
system in question. Unfortunately, this terminology has the potential
to cause confusion, it seems to me, which is why `ignorance' has been
used, more or less as a synonym, in this article. Semantically, it
might seem more sensible to regard knowledge that is \emph{possessed}
as information rather than that which is not possessed. In common
language, we \emph{acquire} information when reducing the uncertainty
of a situation, while according to the technical interpretation of
the word one would say that the system information has been \emph{reduced}.
The terminology is well established, however, and a rebranding of
information theory as ignorance theory would probably not convey a
very positive impression, so we shall have to live with it! At least
the letter $I$ in $S_{I}$ can stand for both ignorance and information
(and `incertainty' as well, if we are inclined to use an archaic word).

The probabilities that the system should take various microstates
evolve in time, and we might hope to be able to characterise this
behaviour, and thereby to model the development of Shannon entropy
and hence our uncertainty. We shall return to this matter in Section
\ref{sec:Stochastic-thermodynamics-and}, but we first investigate
how we might avoid the supposed natural increase in ignorance through
invoking a process of measurement and feedback. It is time to introduce
the demon.

\section{The purpose of the demon\label{sec:The-purpose-of}}

We often study the world in order to exploit its richness for some
useful purpose, or perhaps to allow us to control future events. We
measure the properties of a particular set of materials, for example,
probing at scales that would normally be inaccessible, and then we
work out how to make an object with useful thermal, optical, electrical,
mechanical or chemical properties.

This conforms to the traditional procedure of (a) measurement, (b)
formation of a view on how the system behaves, and (c) exploitation
for a desired outcome. These activities are precisely those of Maxwell's
demon, the infamous imaginary character introduced by James Clerk
Maxwell (and given the evocative name later by Kelvin) as a contribution
to the early discussion of the concept of irreversibility in thermodynamics
and statistical physics \cite{Maxwell71,Leff03}. The demon's particular
skill lies in breaking the second law of thermodynamics, at least
so it would seem, thereby reducing the total entropy of the world
or universe and causing consternation and endless fascination amongst
scientists ever since.

It will become apparent that the viewpoint in this article is that
the demon is a very ordinary creature indeed, performing operations
that are not particularly unusual. He is just a tiny version of one
of us, and indeed it appears this was exactly Maxwell's purpose in
inventing him \cite{Earman98,Earman99}.

Nevertheless, a lengthy discussion of the demon has developed in the
literature, often addressing whether or not the demon's activities
are `illegal' in some sense, or `costly' in another \cite{Leff03,Szilard29,Brillouin51,Landauer61,Bennett73,Bennett82,Bennett03,Earman98,Earman99,HemShen12,Norton05}.
But some recent contributions have taken the form of the construction
and employment of real devices that seem to resemble a demon \cite{Toyabe10,Koski14,Koski14b},
and this has spurred further interest. We will come to these in Section
\ref{sub:Real-experiments}, but first we should discuss the thought
experiments.

\begin{figure}
\begin{centering}
\includegraphics[width=1\columnwidth]{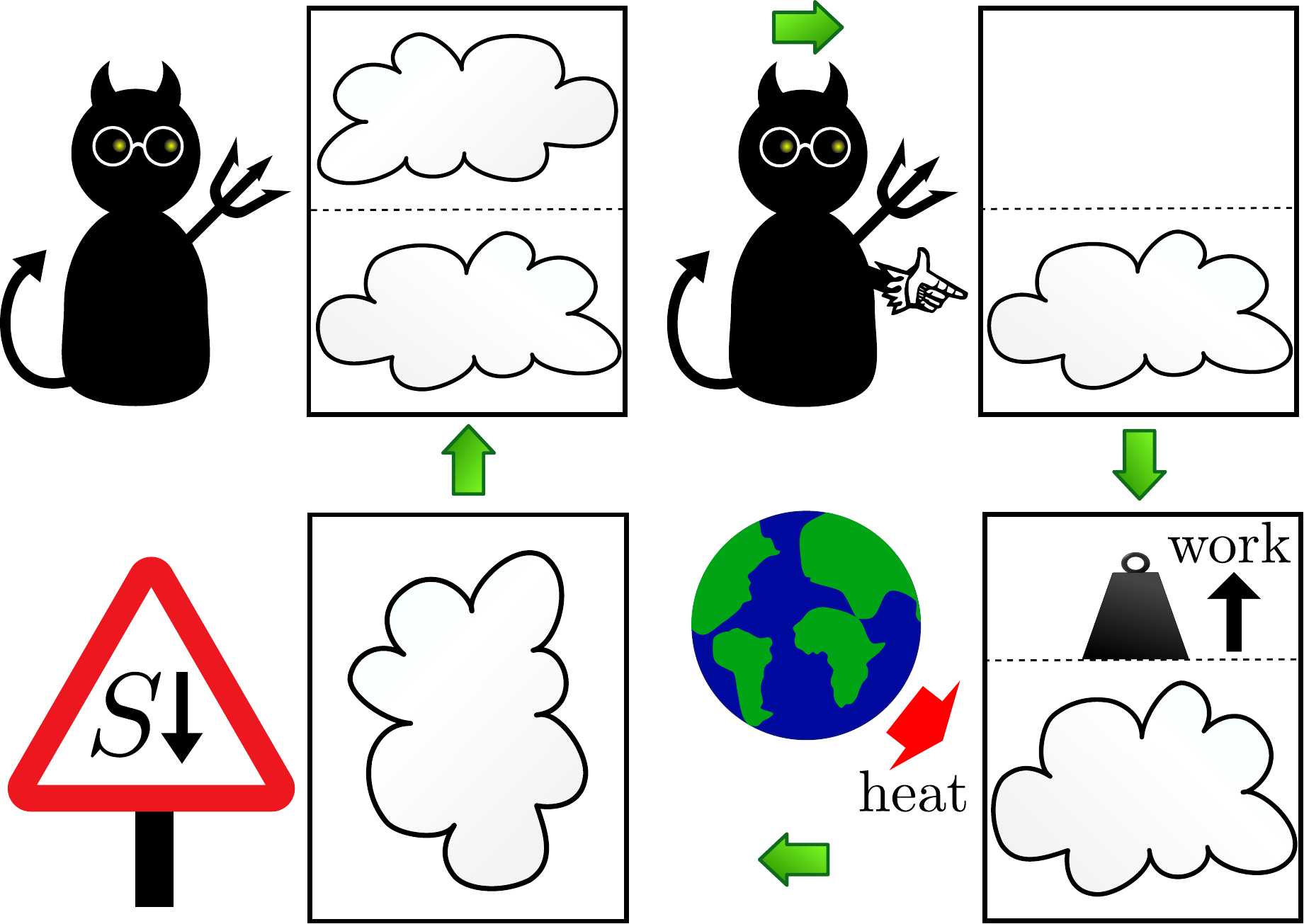}
\par\end{centering}

\caption{Demon at work. He makes a measurement of a system (determining which
half of a box contains an enclosed particle) which allows us to exploit
the reduced state of ignorance to lift a weight, taking energy in
the form of heat from the surrounding environment. The process of
such a Szilard engine \cite{Szilard29} is cyclic, with a guaranteed
decrease in total entropy, as long as we do not probe too deeply into
the effect of the operation on the demon. \label{fig:demon-at-work}}
\end{figure}

\subsection{Thought experiments\label{sub:Thought-experiments}}

The original demon was invested with nimble fingers that would enable
him to sort the molecules of a gas according to their speed. The traditional
picture involves a container separated into two parts, with the demon
able to open and shut a trapdoor in the partition that lies between
them. The demon observes the system and decides to let fast molecules
through the trapdoor in one direction, and slow molecules in the other
direction, so that over the course of time the sorting takes place
\cite{Leff03}.

On the face of it, these operations seem rather unobjectionable, but
they have a particular undesirable consequence. The gas on one side
of the partition after sorting is hotter than the gas on the other
side. So why not then drive a heat engine using this pair of hot and
cold reservoirs? We could exploit the flow of heat to extract some
work, using it to raise a weight (a potential energy store), until
the two gases have come into thermal equilibrium again, presumably
at a cooler temperature than they were to start with. The initial
temperature could be restored by reheating the gas from a suitable
environment, or heat bath, and then we could allow the demon to repeat
the process. Overall, kinetic energy in the form of heat in the bath
would be converted into potential energy stored in the raised weight.
Is this a problem?

The issue is that this would break Kelvin's formulation of the second
law, which specifically declares such a conversion to be impossible
on empirical evidence, at least at the macroscale \cite{fordbook}.
More technically, it would be a system that would automatically evolve
in a way that would reduce our ignorance regarding the exact microstate
of the world.

What does this last statement mean? Well, we are by definition entirely
ignorant of the way energy is held by the constituent parts of a heat
bath. In contrast, moving the height of a weight does not involve
a change in the uncertainty of the way energy is held, only a shift
in the position of the weight. Moving energy from the heat bath to
the potential energy store is therefore accompanied by a reduction
in uncertainty: we have become less ignorant. As we remarked earlier,
a system that moves towards an attractor as time progresses is certainly
feasible dynamically, but our practical experience of the behaviour
of gases and effect of thermodynamic operations makes it hard to accept
that this should apply here. What exactly has the demon done?

There are other scenarios where a demon can engineer a breakage of
the second law. A famous example is the Szilard heat engine, based
on a single particle in a cavity and coupled to a heat bath \cite{Szilard29}.
Without a demon to tell us, we are ignorant as to the position of
the particle in the cavity. We suddenly insert a partition and divide
the cavity into two. One subcavity now contains the particle and the
other does not. We ask the demon which subcavity holds the particle,
as illustrated in the transition from the top left to the top right
image in Figure \ref{fig:demon-at-work}, and then we use that subcavity
as a source of pressure, expanding it to raise a weight (lower right).
The potential energy transferred to the weight comes from heat donated
by the environment. We are now back where we started (lower left)
and we can repeat the operation, through which we reduce our ignorance
of the microstate of the world and break the second law.

Note that we act upon guidance given to us by the demon. We are careful
to specify the demon as the only source of knowledge about the location
of the particle: we discount observation by any other route. For example,
we cannot ourselves touch the partition to gauge the direction from
which the particle collisions are coming, or simply look inside each
cavity. This is the job of the demon. By specifying that the flow
of knowledge passes through the demon, and stating exactly how this
is to be achieved, we can pin down the manner in which our ignorance
is managed.

So, all that we demand is that the demon makes a measurement and tells
us the result. The second law is violated. What could possibly be
the problem with this?

There is a line of argument that somehow the second law has been preserved
through the very act of making the measurement. The literature contains
a history of lively illustrations of how observation might result
in entropy production. For example, the demon might use a torch to
illuminate each subcavity with just enough radiation to determine
the location of the particle \cite{Brillouin51,Gabor64}. Energy is
essentially taken from a potential energy store such as a battery
and converted into photons and ultimately into heat, and so there
will be a payment made in entropy production or the increase in uncertainty
before any work extraction can be performed \cite{Sagawa09,Granger11,Jacobs12,Um15}.

However, the resolution that seems to have attracted the greatest
attention is more abstract than this, and was developed by Bennett
and others \cite{Bennett73,Bennett82,Bennett03,Plenio01} after it
was argued that the act of measurement might in principle not involve
the dissipation of energy and generation of entropy. If that were
possible, how can we save the second law?

I personally am not persuaded by the following resolution, for reasons
to be discussed, but nevertheless here it is. The idea is that after
the demon has made a measurement and informed us of the result, he
needs to be returned to his starting condition in order to make further
measurements. The usual picture is that the demon can take three microstates:
one denoted \emph{ready}, and then two outcome configurations labelled
\emph{left} and \emph{right, }referring to the position of the particle
(top and bottom in Figure \ref{fig:demon-at-work}). The microstate
of the demon after the process of measurement is either \emph{left}
or \emph{right}, and he needs to be converted\emph{ }back into \emph{ready. }

Now, the Landauer principle claims that an operation where the number
of possible configurations of a system is reduced requires work to
be dissipated as heat \cite{Landauer61,Plenio01}, a phenomenon associated
with entropy production, though the foundations of this principle
have been challenged \cite{Norton05,Norton11,Dillenschneider09} as
well as defended \cite{Bub01}. Assuming the principle holds, the
resetting would therefore generate entropy, and analysis shows that
it is never less than the reduction in entropy associated with the
earlier exploitation of the measurement. The second law would be saved!

But the unsatisfactory element is the following. There is nothing
to stop us putting the `used' demon to one side and employing a new
demon, in the \emph{ready} state, to perform the next measurement.
We can repeatedly reduce the entropy of the world and stack up a legion
of demons in their states of \emph{left} and \emph{right}. If we decide
never to reset them, we have apparently broken the second law. It
might be claimed that the law only holds for a cyclic process, the
completion of which would include the resetting of the demons, represented
as the wiping of their memories. It might also be claimed that the
legion of demons has acted as a second heat bath in the process, a
repository of entropy, a possibility that we consider shortly. Nevertheless
the procedure just described allows us to convert energy from a single
heat bath into work endlessly, which is a situation that might still
be regarded as illegal.

The issue is somewhat philosophical, and it is whether a law restricting
the conversion of heat into work, that relies on an eventual reckoning
of the accounts at some unspecified future date, is a law in any real
sense. If a law exists, then it surely cannot offer indefinite periods
of grace. We are essentially being offered a perpetual loan from the
Bank of Negative Entropy!

It is hard to say whether this particular aspect is widely regarded
as unsatisfactory, but it is certainly the case that discussion of
the demon has continued well after the presentation of the argument
of demonic memory erasure. It is important to establish a firm foundation
for a law if it is to have real meaning.

The following line of argument could clarify the situation. The `used'
demons actually represent a store of entropy, corresponding to our
ignorance or uncertainty about their state. They started out \emph{ready}
and end up in either \emph{left} or \emph{right}. Perhaps the resolution
of the issue is that the act of measurement effectively transfers
uncertainty from the system to the demon in a sequence of operations.
The reduced uncertainty of the \emph{system} can then be exploited
to convert heat into work, but if we take into account the earlier
transfer to the demons this will not break a second law, as long as
it is framed as a statement about never-decreasing uncertainty \cite{Plenio01}.

This sounds attractive, but it cannot be the whole story since the
act of measurement can sometimes involve a \emph{reduction} instead
of an increase in our ignorance of the state of the demon, as we shall
see in Section \ref{sec:A-simple-model}. Moreover, a rather particular
concept of a demon has been employed here, and we might wonder whether
clarity about the initial \emph{ready} state is absolutely necessary.
Why not start out in \emph{left} or \emph{right}? Furthermore, although
the nature of the dynamics of measurement has not been made explicit,
the implication is that the demon and system are isolated from the
rest of the world. Measurements are not always made in these circumstances.

As touched upon earlier, the demon debate continues, in part, because
it has been hard to pin down what entropy actually is, and what the
second law exactly requires. It is perhaps best to avoid discussions
that do not frame the matter in a proper dynamical context. We should
be cautious in making imprecise references to `measurement' and the
sometimes counterintuitive notion of `information' in this context.
These are issues that are addressed by stochastic thermodynamics,
as well as related treatments that use deterministic dynamics \cite{Lebowitz99,Wang02,Evans02,Ford15c},
as we shall see in Section \ref{sec:Stochastic-thermodynamics-and}.
We shall return to these topics in Section \ref{sec:The-management-of},
but we first briefly discuss some recent experimental investigations
of demonic behaviour.

\subsection{Real experiments\label{sub:Real-experiments}}

The claim that the demon is just a tiny `one of us' suggests that
we should be able to perform experiments to demonstrate his actions.
These take the form of feedback processes acting on systems with few
degrees of freedom and subject to environmental noise. Various scenarios
that involve mechanical, electrical or chemical processes have been
made possible through recent advances in nanotechnology and several
have been reviewed \cite{Parrondo15,Pekola15,Vinjanampathy15,Lutz15}.

At present, such experiments are not able to challenge the second
law since the measurement procedures are by no means ideal. They require
the dissipation of considerable amounts of externally stored potential
energy. The focus of attention has instead been on the exploitation
of the data acquired by a measuring device: that is, its use in controlling
the manipulation of the system to some advantage. This has been referred
to as the conversion of `information' into free energy.

\begin{figure}
\begin{centering}
\includegraphics[width=1\columnwidth]{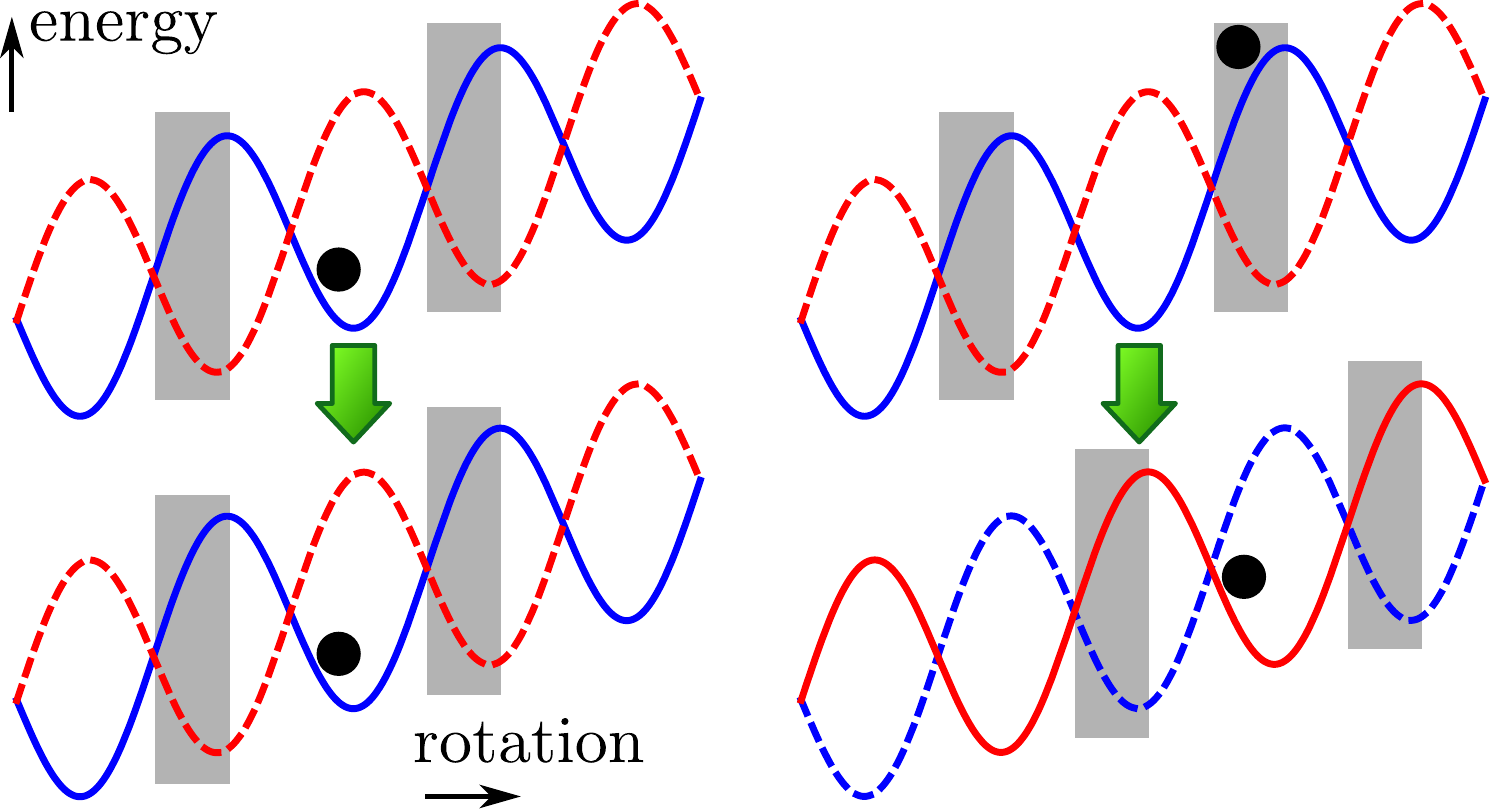}
\par\end{centering}

\caption{The exploitation of a measurement in the experiment by Toyabe \emph{et
}al\emph{. }\cite{Toyabe10}\emph{. }If a particle is found in the
grey zones, as shown in the situation on the right, the external potential
is switched from the solid to the dashed profile; otherwise no action
is taken, as shown on the left. Thermal fluctuations are more likely
to take the particle into the grey zone on the right hand side of
each local well, and the feedback therefore promotes a walk up the
`staircase' towards the right, producing rotational motion that arises
from the transfer of heat from the environment. This is an illustration
of part of the operation of an autonomous Maxwell's demon, though
no account is taken of the potential energy dissipation inherent in
the act of measurement. (Figure adapted from Toyabe \emph{et }al\emph{.}
\cite{Toyabe10}, with permission). \label{fig:expts}}
\end{figure}

Toyabe \emph{et al.} \cite{Toyabe10} presented a colloidal system
that could be made to rotate, and essentially acquire additional potential
energy, under the influence of a heat bath together with force fields
externally controlled by measurement feedback. The operation of the
system is summarised in Figure \ref{fig:expts}, where it is characterised
as the motion of a particle over an endless profile of potential energy
as a function of rotational angle. The feedback system is designed
to make changes in the potential energy surface in a way that tends
to preserve a random fluctuation of the particle in one rotational
direction, analogous to the demon's opening of a trapdoor when a gas
particle with the right velocity comes into view. The demonstration
of demonic activity is not quite complete, however, since there is
potential energy dissipation in the act of measurement as well as
in the practical matter of switching the potential. Nevertheless,
the experiment demonstrates the principle of feedback manipulation
and some of its thermodynamic consequences.

Similar demonstrations of the use of feedback from measurement in
an electronic system have been presented by Koski \emph{et al.} using
a single electron box \cite{Koski14,Koski14b,Pekola15}. The system
can be regarded effectively as a particle that takes one of two degenerate
microstates, while coupled to an environment that introduces uncertainty
into the situation. The degeneracy can be broken by manipulating an
external field, lifting the energy of one of the microstates up while
the other is reduced. By determining which of the microstates is actually
taken by the particle, the manipulation can be tailored to reduce
the potential energy of the system, transferring the difference to
a store. Slowly returning the field to its original value allows the
system to recover the lost energy through the absorption of heat from
the environment, and the cycle can then be repeated. The key practical
matter is the identification of the actual microstate and the implementation
of the exploitation at sufficient speed, and the experiment demonstrates
that this is achievable. Many cycles can be carried out and the statistics
of the operation are very precise. Once again, the focus of attention
is on demonstrating the thermodynamic benefits of feedback and not
on whether the second law has been challenged.

B\'erut \emph{et al.} \cite{Berut12} considered a different aspect
of the demon narrative by demonstrating that a minimum amount of external
mechanical work has to be performed (and dissipated as heat) in order
to reset or convert a two-state memory of uncertain configuration
into one of definite configuration. They successfully demonstrated
that a minimum of $kT\ln2$ of heat production was required per reset
operation, the Landauer limit \cite{Landauer61}.  In the experiment,
a colloidal particle held on one side or other of a double potential
well, generated optically, was manipulated by an external force field
such that at the end of the operation, it definitely occupied one
side. The strength of the field and the trajectory of the particle
could be used to determine the external work provided, and the average
over many realisations, obtained for a range of process times, always
exceeded the Landauer limit. The consequence of reducing uncertainty
in the memory is the passage of heat into the environment. Jun \emph{et
al.} \cite{Jun14} reported a similar experimental demonstration of
Landauer's principle.

We also mention studies of systems that exhibit self-sorting behaviour,
a property that we might either regard as an example of a successful
demon at work, or as outside the remit of thermodynamics. The concept
of a diode that allows the passage of atoms in one direction only
has been demonstrated with an optical system \cite{Raizen05,Ruschhaupt04}:
it can be used to compress atoms into a smaller volume without apparent
expenditure of work. It can also be exploited to extract heat from
a gas \cite{Raizen09}, though not necessarily with its conversion
entirely into work.

As nanotechnology develops, further demonstrations of manipulation
at molecular scales, taking advantage of measurement, will follow.
We now turn our attention to the thermodynamic framework that seems
to be the most appropriate when we carry out processes at this level.

\section{Stochastic thermodynamics and the second law\label{sec:Stochastic-thermodynamics-and}}

Stochastic thermodynamics is based on stochastic equations of motion
describing the evolution of a system, with noise representing interactions
with an environment. This is a \emph{model} of the world: it is not
reality, which presumably involves the deterministic evolution of
the system and environment together. Nevertheless such a model can
capture the behaviour of the system we wish to represent, namely dissipative
(energy sharing) in character, subject to fluctuations and lacking
predictability. It is not a viewpoint that resolves the old paradox
about how time-irreversible phenomena can arise from time-reversible
fundamental dynamics \cite{Price97,North11,Werndl12}. The arrow of
time is inserted by hand, in the sense that the dynamics are intended
to account for evolution \emph{forward} in time starting from some
initial condition, but not necessarily backwards.

As an illustration of such stochastic differential equations (SDEs),
we consider
\begin{equation}
\frac{dv}{dt}=-\gamma v+\frac{F(x,t)}{m}+\left(\frac{2kT(t)\gamma}{m}\right)^{1/2}\xi(t),\label{eq:1}
\end{equation}
with $v=dx/dt$, where $x$ and $v$ are the position and velocity
of a particle, respectively, $t$ is time, $\gamma$ is a friction
coefficient, $F(x,t)$ is a force field acting on the particle, $m$
is the particle mass, $k$ is Boltzmann's constant, $T(t)$ is the
temperature of the environment, and $\xi(t)$ is a random `white'
noise with statistical properties $\langle\xi(t)\rangle=0$ and $\langle\xi(t)\xi(t^{\prime})\rangle=\delta(t-t^{\prime})$
\cite{Risken89,vanKampen07,Gardiner09}, where the brackets represent
an average over all possible values of the noise $\xi$. The second
of these conditions implies that the noise at different times is uncorrelated,
or lacking memory.

\begin{figure}
\begin{centering}
\includegraphics[width=1\columnwidth]{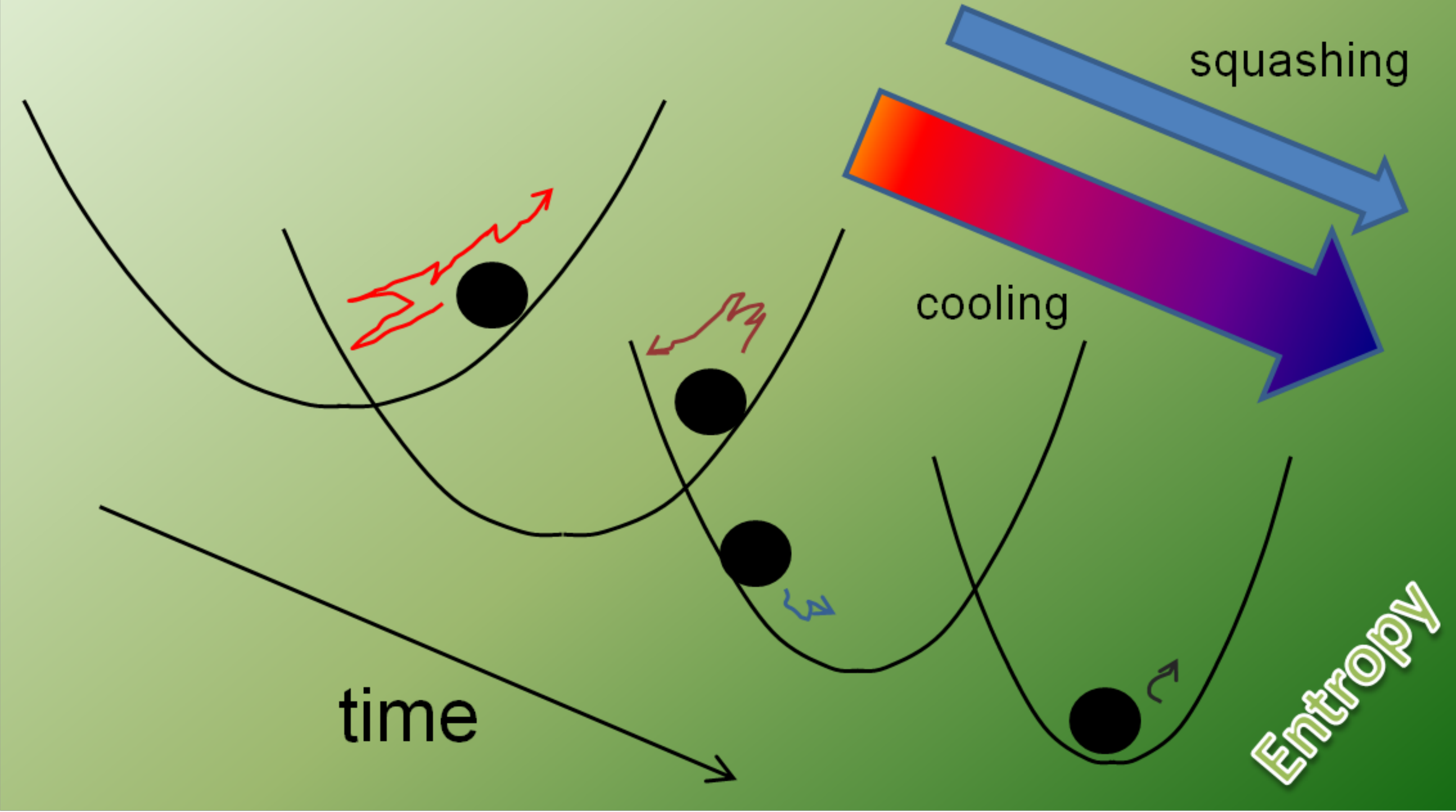}
\par\end{centering}

\caption{The stochastic dynamics of a particle under the influence of a time-dependent
potential (illustrating a work process such as mechanical squashing)
and a random force or noise with a time-dependent strength (corresponding
here to a reduction in environmental temperature, driving cooling).
The effect of the stochastic thermodynamics is illustrated by the
deepening colouration of the environment: in a manner of speaking
the thermomechanical processing of the particle on average darkens
the world with the production of stochastic entropy.\label{fig:1}}
\end{figure}

The dynamics describe a one dimensional Brownian motion, and are designed
to relax the system to canonical equilibrium, as long as the force
is related to a potential $\phi$ and the temperature is constant,
such that the eventual probability density function (pdf) over $x$
and $v$ is $p_{{\rm eq}}\propto\exp[-(\phi+mv^{2}/2)/kT]$. A variety
of more complicated stochastic equations of motion can be imagined,
incorporating noise with memory, or a more elaborate friction term,
but this example will serve to illustrate the ideas. The evolution
of the system is illustrated in Figure \ref{fig:1}.

The second requirement of stochastic thermodynamics is a definition
of the entropy production associated with a possible realisation of
the motion \cite{sekimoto1,sekimoto2,seifertoriginal,seifertprinciples}.
This is fundamentally a measure of the probabilistic mechanical irreversibility
of the motion. For a time interval $0\le t\le\tau$, the dynamics
can generate a trajectory $\vec{\boldsymbol{x}},\vec{\boldsymbol{v}}$
(where $\vec{\boldsymbol{x}}$ represents a function $x(t)$ in the
specified time interval) according to a probability density function
${\cal P}[\vec{\boldsymbol{x}},\vec{\boldsymbol{v}}]$.

The dynamics are also capable of generating an \emph{antitrajectory}
$\vec{\boldsymbol{x}}^{\dagger},\vec{\boldsymbol{v}}^{\dagger}$ in
the period $\tau\le t\le2\tau$, following an inversion of the velocity
at time $\tau$, where $x^{\dagger}(t)=x(2\tau-t)$ and $v^{\dagger}(t)=-v(2\tau-t)$.
The antitrajectory starts at $x(\tau),-v(\tau)$ and ends at $x(0),-v(0)$,
driven by a reversed time evolution of the force field and environmental
temperature \cite{fordbook,SpinneyFordChapter12,Ford15c}: it is the
`time-reversed' partner of $\vec{\boldsymbol{x}},\vec{\boldsymbol{v}}$
\cite{SpinneyFord12a,SpinneyFord12b,SpinneyFord12c}, though to be
absolutely clear we do \emph{not} consider evolution of the system
into the past.

We denote the probability density that an antitrajectory is generated
in the period $\tau\le t\le2\tau$ as ${\cal P}^{{\rm R}}[\vec{\boldsymbol{x}}^{\dagger},\vec{\boldsymbol{v}}^{\dagger}]$,
and the total entropy production associated with the trajectory $\vec{\boldsymbol{x}},\vec{\boldsymbol{v}}$
is then defined by
\begin{equation}
\Delta s_{{\rm tot}}[\vec{\boldsymbol{x}},\vec{\boldsymbol{v}}]=\ln\left[\frac{{\cal P}[\vec{\boldsymbol{x}},\vec{\boldsymbol{v}}]}{{\cal P}^{{\rm R}}[\vec{\boldsymbol{x}}^{\dagger},\vec{\boldsymbol{v}}^{\dagger}]}\right].\label{eq:2}
\end{equation}
After multiplication by Boltzmann's constant and averaging over all
realisations of the motion, this corresponds to the production of
thermodynamic entropy in the process. In a condition of thermal equilibrium,
defined to be a situation where the dynamics generate a trajectory
and its time-reversed partner with equal likelihood, the entropy production
associated with \emph{all} feasible trajectories will vanish.

An increment in $\Delta s_{{\rm tot}}$ for the specified dynamics
may be shown to be given by \cite{seifertoriginal,SpinneyFord12b}
\begin{equation}
d\Delta s_{{\rm tot}}=-d[\ln p(x,v,t)]-\frac{1}{kT(t)}d\left(\frac{mv^{2}}{2}\right)+\frac{F(x,t)}{kT(t)}dx,\label{eq:3}
\end{equation}
which is an expression with great intuitive value. The second term
is the negative increment in the kinetic energy of the particle over
the time interval $dt$, and the third term is the negative increment
in its potential energy, both divided by the environmental temperature.
They represent a positive increment in the energy of the environment
(a heat transfer $dQ_{{\rm env}}$) divided by the temperature, corresponding
to a Clausius-type incremental change $d\Delta s_{{\rm env}}=dQ_{{\rm env}}/kT$
in the entropy of the environment in the interval of time $dt$.

Seifert defined a stochastic system entropy $s_{{\rm sys}}=-\ln p(x,v,t)$
in terms of the evolving phase space probability density function
$p$ generated by the stochastic dynamics \cite{seifertoriginal},
such that we can write
\begin{equation}
d\Delta s_{{\rm tot}}=d\Delta s_{{\rm sys}}+d\Delta s_{{\rm env}}.\label{eq:4}
\end{equation}
The evaluation of $\Delta s_{{\rm tot}}$ for a specific realisation
of the motion clearly requires us to determine the evolution of the
pdf $p$, as well as the system variables $x$ and $v$, for which
we need to solve a Fokker-Planck equation \cite{Risken89}.  Using
the evolving system pdf to study changes in the Shannon entropy of
the system is itself a form of stochastic thermodynamics \cite{Tome15},
but the formulation based on the definition of a stochastic entropy
production arguably goes deeper, since it allows us to address the
production of entropy in both system and environment and especially
to recognise that fluctuations in these quantities can occur.

We now come to a key point: the total stochastic entropy production
satisfies the following \emph{integral fluctuation relation} \cite{seifertoriginal}
\begin{equation}
\langle\exp(-\Delta s_{{\rm tot}})\rangle=1,\label{eq:5}
\end{equation}
which leads immediately to $\langle\Delta s_{{\rm tot}}\rangle\ge0$
and hence $d\langle\Delta s_{{\rm tot}}\rangle\ge0$, where the angled
brackets now denote an average over all possible trajectories taken
by the system. These inequalities may be regarded as an expression
of the second law of thermodynamics in this framework. The increment
in thermodynamic entropy production $dS_{{\rm tot}}$ over the period
$dt$ is taken to be $d\langle\Delta s_{{\rm tot}}\rangle$, the average
of all possible increments in the total stochastic entropy production
in that interval. The limit with $\langle\Delta s_{{\rm tot}}\rangle=0$
can be achieved by performing the process exceedingly slowly, or quasistatically
\cite{SpinneyFordChapter12}. This would be a reversible process;
all others are then irreversible.

Furthermore, the integral fluctuation relation leads to the celebrated
Jarzynski equality \cite{jarzynski1996nonequilibrium}, which we shall
be using later on in a discussion of the processes of measurement
and exploitation through the demon. If a system starts out in canonical
equilibrium, and is then subjected to time-evolving Hamiltonian forces
while the environment remains at a constant temperature, it can be
shown that
\begin{equation}
\langle\exp(-W/kT)\rangle=\exp(-\Delta F/kT),\label{eq:6}
\end{equation}
where $W$ is the mechanical work performed on the system in such
a process, a quantity that depends on the trajectory taken, and $\Delta F$
is the change in Helmholtz free energy corresponding to the change
in Hamiltonian.

A further important conclusion is that, while the average increment
in total stochastic entropy production is positive or zero, the average
increments in the stochastic entropy production in the system and
environment can be negative. For a process that takes place over an
interval $\tau$, we can compute the difference between the final
and initial averages of $s_{{\rm sys}}$ to determine the average
change in the stochastic entropy of the system, namely
\begin{eqnarray}
 &  & \langle\Delta s_{{\rm sys}}\rangle=\Delta\langle s_{{\rm sys}}\rangle=-\!\int\!p(x_{\tau},v_{\tau},\tau)\ln p(x_{\tau},v_{\tau},\tau)dx_{\tau}dv_{\tau}\nonumber \\
 &  & \qquad\quad+\int p(x_{0},v_{0},0)\ln p(x_{0},v_{0},0)dx_{0}dv_{0},\label{eq:7}
\end{eqnarray}
where $x_{t}=x(t)$ and $v_{t}=v(t)$, so that $\langle\Delta s_{{\rm sys}}\rangle=S_{I}(\tau)-S_{I}(0)$.

We regard the Shannon entropy to be a measure of our ignorance or
uncertainty of the system microstate, but this does \emph{not} have
to increase as time progresses: it is ignorance of the microstate
of the system plus environment that is obliged never to decrease.
This allows us to imagine operations, possibly involving the injection
of potential energy from some external store, that can reduce our
uncertainty of the system microstate. These might correspond to processes
of measurement. The uncertainty must be taken up by other components
of the world though, and this leads us to consider next how an exchange
of ignorance, and potentially an exploitation of measurements, can
be managed against a backdrop of the likely continued generation of
uncertainty.

\section{The management of ignorance under stochastic dynamics \label{sec:The-management-of}}

If stochastic thermodynamics tells us that our ignorance of the world
can be shifted about between its component parts, we might be able
to discuss measurement and the exploitation of acquired knowledge
within such a framework; in other words, to analyse the activities
of a demon.

Let us first consider a particular scenario of ignorance management
that will correspond to the Bennett resolution of the action of Maxwell's
demon. Imagine that the world consists of a system, a demon, a store
for potential energy, and with everything else represented by an environment.
The coupling and uncoupling between the demon and the system, and
the store and the system, can be programmed as we wish. Imagine that
the demon is coupled to the system in such a way that uncertainty
in the state of the system is reduced while that of the demon is raised.
The resulting improved clarity about the state of the system can then
be exploited in the manner considered in the thought experiments of
Section \ref{sub:Thought-experiments}. Heat energy from the environment
may thereby be converted into potential energy in the store, in principle
reversibly, i.e. without an overall increase in uncertainty. If we
neglected to consider how to return the demon to his original state,
we would be led to believe that the demon's activities had broken
the second law.

The process will only have been made possible by the transfer of uncertainty
to the demon. The situation is resolved by resetting the demon, which
involves the transfer of uncertainty from him back to the environment,
together with some additional generation if this is not done quasistatically.
Landauer's argument is that this process is associated with the dissipation
of potential energy as heat and thus the energy store is deprived
of its earlier gains. There is no violation of the second law if it
is regarded as a statement that total uncertainty should never decrease
with time; in this viewpoint it is only incidentally associated with
the conversion of potential energy into heat and vice versa. Nevertheless,
in this scenario it seems that we can convert heat from the environment
into work and, as a by-product, simply pile up demons of an uncertain
state. Can we accept such a loan from the Bank of Negative Entropy?

\begin{figure}
\begin{centering}
\includegraphics[width=1\columnwidth]{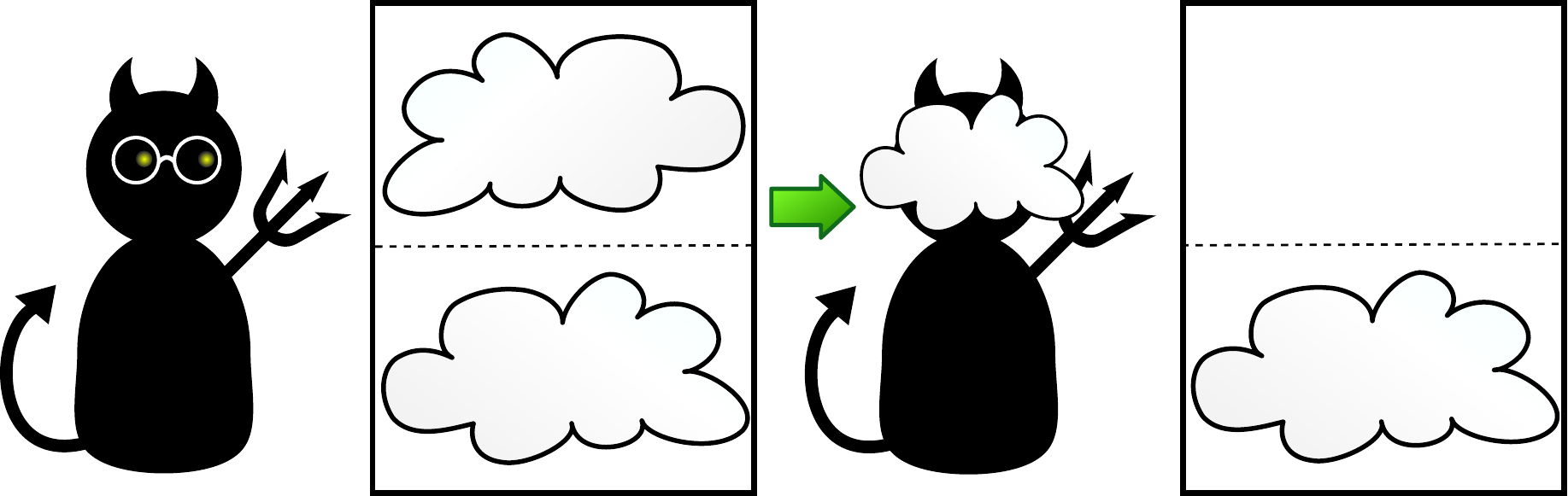}
\par\end{centering}

\caption{A scenario where a measurement is performed by transferring uncertainty
from the system to the demon. The system might then be returned to
its initial state with an associated conversion of environmental heat
into stored potential energy, and reduction in entropy. In order to
repeat the operation a fresh demon is needed: ultimately the `used'
demons have to be reset, which requires the expenditure of work and
the generation of entropy. This is essentially the Bennett rationalisation
of the action of a demon \cite{Bennett73}. \label{fig:Confused-demon}}
\end{figure}

This scenario is illustrated in Figure \ref{fig:Confused-demon}.
It has some appealing aspects, but it is not necessarily an appropriate
description of events that take place between system, demon and environment
in a framework of stochastic thermodynamics. Using a simple model
in Section \ref{sec:A-simple-model}, we will show that the measurement
procedure can render both the system and demon in a \emph{reduced}
state of uncertainty, while there is an increase in uncertainty in
the environment to compensate. The latter is associated with a flow
of heat, transferred from a potential energy store, and a measurement
is therefore paid for through the \emph{upfront} performance of work
\cite{Sagawa09}. This contrasts with the viewpoint where work is
performed at a later time to reset the demons and repay the earlier
loan of entropy reduction. The process where heat to work conversion
can take place at the cost of piling up used demons simply does not
arise. The transformation of heat into work is not achieved solely
by transferring uncertainty: we pay for it in prior dissipative work.

We can understand how this new scenario emerges by reflecting on what
precisely constitutes a measurement by the demon. We must avoid too
great a level of abstraction and should imagine how the measurement
and exploitation are to be represented and implemented through the
dynamics. In order not to run the risk of falling into confusion we
must not refer in loose terms to the acquisition of knowledge and
the taking of appropriate action. We must set all our considerations
within a practical, dynamical framework.

Let us now consider a key point, which is that the act of measurement
of the system by the demon is the process of becoming dynamically
correlated with it. This requires a coupling to the energy store,
and could take place while in contact with the environment. The microstate
of the demon, when correlated, can serve as a proxy for the microstate
of the system and further coupling of the system to the energy store
can be programmed in such a way as to exploit the situation. The state
of the demon after the measurement thereby brings about a suitably
designed set of actions.

But we might imagine that the programme of exploitation could be determined
by the microstate of the system itself, so that we could cut out the
demon entirely. We shall discuss this in some detail in Section \ref{sec:Rules-for-demons},
but let us assert now that systems where the dynamics are self-sorting,
corresponding to the existence of an attractor of some kind, are to
be excluded from consideration. Such systems would potentially not
offer the thermodynamic behaviour that we seek, particularly the stability
of a canonical equilibrium state. We assert that, in order to sort
a system, we need to make a measurement through the intermediary of
the demon, and we require the demon to be decoupled from the system
before the exploitation is implemented, or else the composite demon/system
dynamics would be self-sorting. There are rules, it would seem, for
demons: they are there to provide a feedback mechanism, while never
becoming caught up in the consequences of the feedback. We shall comment
further on these rules in Section \ref{sec:Rules-for-demons}.

\begin{figure}
\begin{centering}
\includegraphics[width=1\columnwidth]{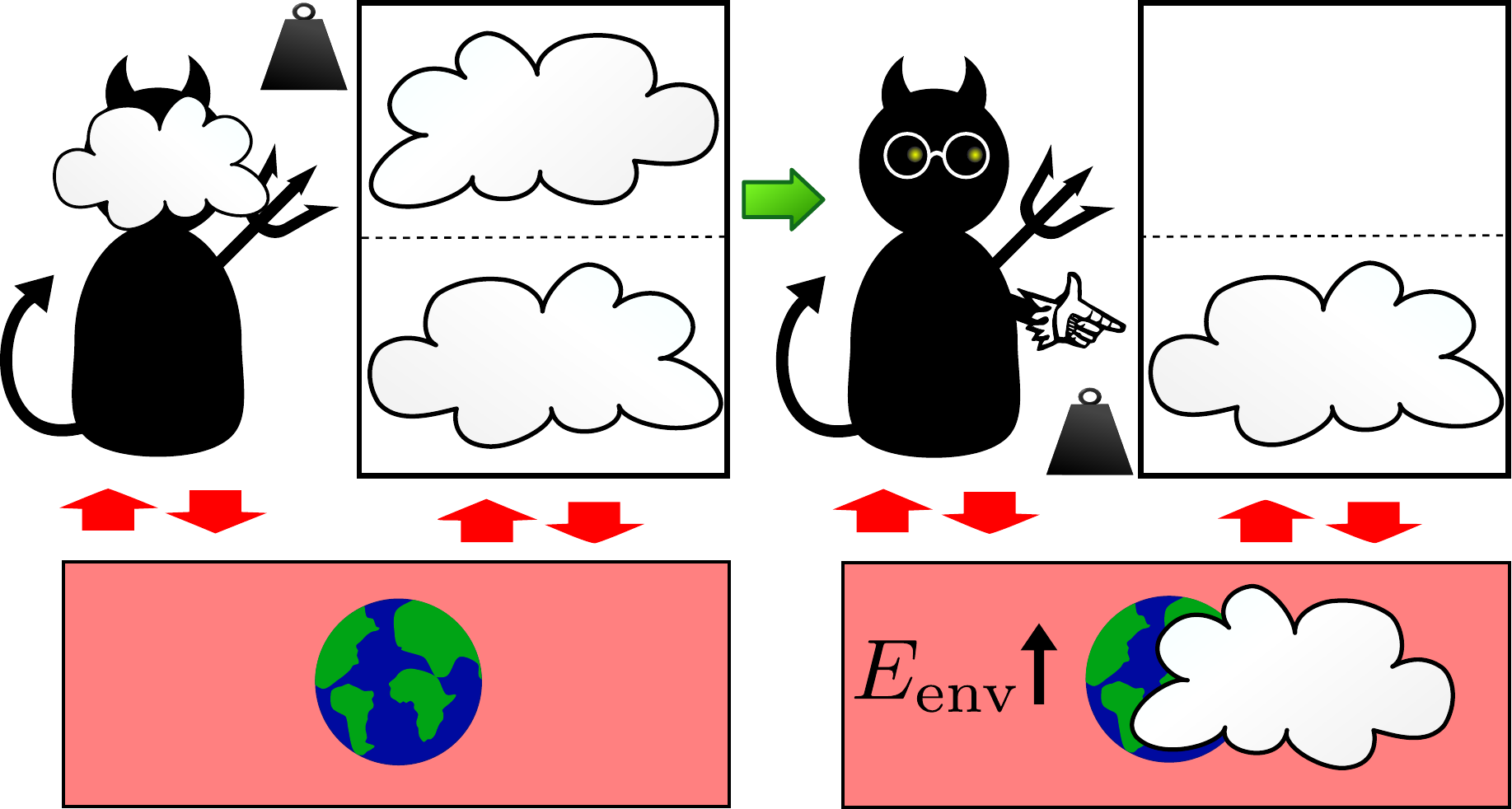}
\par\end{centering}

\caption{A demon performs a measurement within a framework of stochastic thermodynamics.
The investment of potential energy, taken from a store, in the dynamical
coupling of the demon to the system brings about a correlation between
them that persists after decoupling. This results in a reduction in
uncertainty in their microstates compared with an initial situation
of thermal equilibrium with the environment. In the process, however,
potential energy is transferred, on average, to the environment, bringing
about increases in mean environmental energy $E_{{\rm env}}$, in
thermodynamic entropy of the environment, and in the overall uncertainty
of the microstate of the universe. \label{fig:Sharpened-demon}}
\end{figure}

We emphasise that according to an explicit consideration of the dynamics,
the demon, after the measurement has been made according to this scenario,
is typically left in a state of \emph{reduced} uncertainty with respect
to its initial equilibrium state. This is the opposite of the situation
in Figure \ref{fig:Confused-demon}. In order that the demon should
be restored to his original condition, it is sufficient simply to
couple him to the environment, whereupon an irreversible relaxation
process take place, typically involving heat exchange and generating
the required uncertainty. Alternatively the used demon could be employed
for the next measurement: the microstate of the demon prior to coupling
to the system is not crucial. A third option is that the post-measurement
demon might actually be exploited in some way to recover heat from
the environment, to be briefly discussed in Section \ref{sub:Exploitation}.
The scenario is illustrated in Figure \ref{fig:Sharpened-demon},
and details of how it can be realised in practice are given in Section
\ref{sec:A-simple-model}.

Sequences of measurement and exploitation are able, on average, to
convert heat in the environment to mechanical work, a violation of
Kelvin's statement of the second law, but some potential energy has
to be taken from a store and dissipated in order to make the measurement.
Furthermore, and crucially, analysis using stochastic thermodynamics
shows that the average of this work of measurement \emph{exceeds}
the average recovery of heat made possible by exploiting the measurement
\cite{Sagawa09}. This is arguably a much more satisfactory outcome
than the scenario where the dissipation of work to `save' the second
law is to be carried out at some unspecified time in the future. It
is in the tradition of the arguments that were current before the
Bennett resolution. Of course, the second law refers to expected or
averaged behaviour and fluctuations are certainly feasible where the
expenditure associated with measurement is less than the return made
through exploitation. We now consider an analysis that underpins these
claims.

\section{An explicit model of measurement and exploitation\label{sec:A-simple-model}}

The issues discussed in the last section can be illustrated more clearly
using a specific dynamical system and demon or measuring device. A
number of such models have been presented in the literature \cite{Granger11,Abreu11,Mandal12,Mandal13,Barato13,Strasberg13,Maitland15},
and the one we present is similarly simple and partly analytically
tractable. We consider the system to be a 1-d harmonic oscillator,
and the demon/measuring device to be another harmonic oscillator that
can be coupled to, and decoupled from the system through a further
harmonic spring. Both system and demon are affected by noise from
the environment, and the change in coupling is brought about by the
supply of potential energy from a store. Once a correlation has been
established between demon and system, they are decoupled, and the
system is then manipulated in such a way that causes energy to pass
from the environment into the store, informed by the microstate of
the demon. We shall consider each of these processes using an overdamped
version of the relevant stochastic dynamics, essentially Eq. (\ref{eq:1})
with the acceleration $dv/dt$ set to zero.

\begin{figure}
\begin{centering}
\includegraphics[width=1\columnwidth]{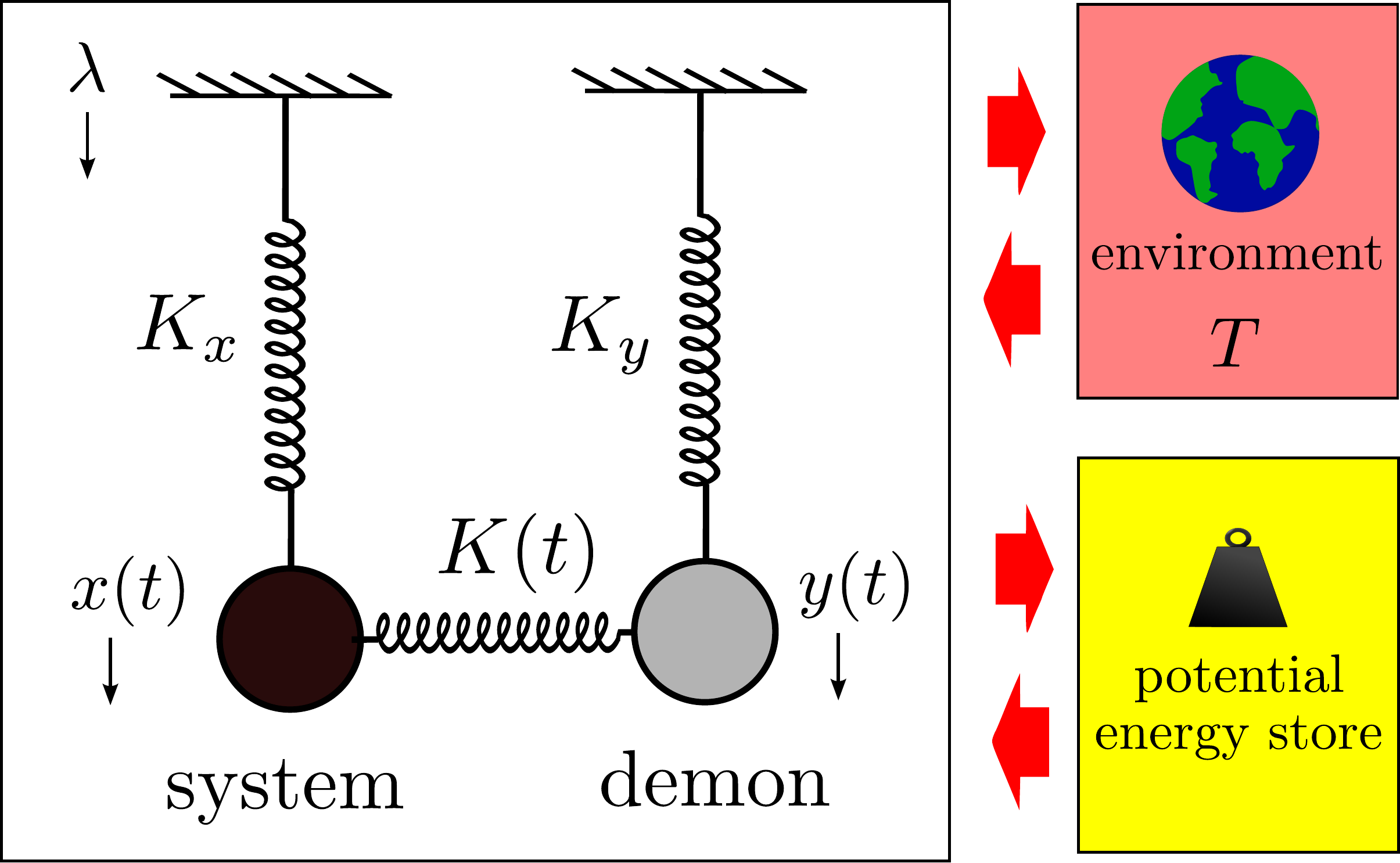}
\par\end{centering}

\centering{}\caption{The system and device are harmonic oscillators that evolve under the
influence of an environmental heat bath at temperature $T$. They
are brought into correlation by a coupling provided by a further spring,
drawing upon a potential energy store. The system and demon displacements,
$x$ and $y$ respectively, evolve according to Eqs. (\ref{eq:8})
and (\ref{eq:9}). \label{fig:system+device}}
\end{figure}

The stochastic differential equations are
\begin{eqnarray}
\!\!\frac{dx}{dt} & \!= & \!-\frac{K{}_{x}}{m\gamma}[x-\lambda]\!-\!\frac{K}{m\gamma}(x-y)\!+\!\left[\frac{2kT}{m\gamma}\right]^{\frac{1}{2}}\!\!\xi_{x}(t),\quad\label{eq:8}\\
\frac{dy}{dt} & = & -\frac{K_{y}}{m^{\prime}\gamma^{\prime}}y-\frac{K}{m^{\prime}\gamma^{\prime}}(y-x)+\left[\frac{2kT}{m^{\prime}\gamma^{\prime}}\right]^{\frac{1}{2}}\xi_{y}(t),\label{eq:9}
\end{eqnarray}
where the displacements of system and demon are given by $x$ and
$y$, respectively. The terms on the right hand side in each equation
represent the intrinsic and coupling spring forces, with strengths
$K_{x}$, $K_{y}$ and $K$ that might depend on time (in doing so
drawing energy from the potential energy store), and white environmental
noise described by independent random variables $\xi_{x}$ and $\xi_{y}$.
The mass and friction coefficient for the demon are $m^{\prime}$
and $\gamma^{\prime}$. There is an additional time-dependent parameter
$\lambda$ that represents the position of the point to which the
system spring is tethered, to be discussed briefly when we consider
exploitation. The system and demon are illustrated in Figure \ref{fig:system+device}.

We consider four intervals of time. In the period $-\infty\le t\le-\tau_{{\rm m}}$
the coupling spring strength $K$ is zero, the system and demon spring
strengths take constant values $\kappa_{x}$ and $\kappa_{y}$, the
system tether position parameter $\lambda$ is zero, and by the end
of the period, a relaxed equilibrium state is established described
by $p_{{\rm eq}}^{x}(x)p_{{\rm eq}}^{y}(y)$ where $p_{{\rm eq}}^{x}(x)=(\kappa_{x}/2\pi kT)^{1/2}\exp[-\kappa_{x}x^{2}/2kT]$
and $p_{{\rm eq}}^{y}(y)=(\kappa_{y}/2\pi kT)^{1/2}\exp[-\kappa_{y}y^{2}/2kT]$.

In the next time interval $-\tau_{{\rm m}}\le t\le0$ the system and
demon are coupled by evolving $K$, with it starting and ending at
zero and drawing upon potential energy from the store. The system
and demon become correlated and the displacement of the demon will
thereafter provide a measurement of the displacement (microstate)
of the system.

In the third period $0\le t\le\tau_{{\rm e}}$ the measurement is
exploited through a sequence of changes in the strength $K_{x}$ of
the system spring and the position $\lambda$ of the tethering point.
It is easy to imagine a process that transfers potential energy from
the system to the store, and then allows the system to absorb heat
from the environment as it relaxes back to equilibrium. For example,
if the demon displacement $y$ is used as an estimate of the system
displacement $x$, the tether position $\lambda$ could be moved to
$y$ in the hope that the system spring can thereby be relaxed to
some extent, harvesting potential energy to the store. Optimal exploitation
sequences were studied by Abreu and Seifert \cite{Abreu11}, but we
need not consider them in detail here.

The role of the demon in this period is to specify the exploitation
sequence through the value of the displacement $y$ at some instant
of time, or perhaps over an extended time interval. Formally, the
parameters $K_{x}(t)$ and $\lambda(t)$ are to be causally determined
by $y(t)$ in the exploitation interval. The nature of the continued
evolution of $y$, however, can take a variety of forms. The demon
could remain coupled to the environment such that it relaxes back
towards an equilibrium state; or the coupling could be removed leaving
the demon in a state of harmonic oscillation; or the displacement
could simply be frozen at some point. In principle, a further option
would be to carry out a procedure to transfer heat from the environment
to the potential energy store, taking advantage of the demon's nonequilibrium
state. But for our purposes, the main objective of the process of
ignorance management by the demon has been accomplished already.

In the final period $\tau_{{\rm e}}\le t\le\infty$, the system, and
possibly the demon, relax back to equilibrium, the force parameters
having been returned to their initial time-independent values. It
is then that we can take stock of the overall transfer of heat from
the environment to potential energy in the store. Let us now consider
the measurement and exploitation phases in greater detail.

\subsection{Measurement\label{sub:Measurement}}

At the start of the measurement interval the system-demon coupling
Hamiltonian $H_{{\rm m}}(x,y,K)=K\,(x-y)^{2}/2$ is equal to zero
and the system and demon are in separate equilibrium states. At the
end of this interval the coupling has returned to zero but the system
and demon will in general be correlated and in a nonequilibrium state.
We assume the system and demon spring strengths $K_{x}$ and $K_{y}$
remain constant throughout measurement so the change in free energy
of the system-demon composite over this period is zero. However, the
work done during measurement, the energy drawn from the potential
energy store, is not zero, being given by
\begin{eqnarray}
W_{{\rm m}} & = & \int_{-\tau_{{\rm m}}}^{0}\frac{\partial H_{{\rm m}}}{\partial K}\frac{dK}{dt}dt\nonumber \\
 & = & \int_{-\tau_{{\rm m}}}^{0}\frac{1}{2}\frac{dK(t)}{dt}[x(t)-y(t)]^{2}dt,\label{eq:10}
\end{eqnarray}
and it satisfies a Jarzynski equality
\begin{equation}
\langle\exp(-W_{{\rm m}}/kT)\rangle=1,\label{eq:11}
\end{equation}
implying that $\langle W_{{\rm m}}\rangle\ge0$, where the bracket
notation refers to an average over all trajectories of $x(t)$ and
$y(t)$ that can take place in the period.

Note that an outcome $\langle W_{{\rm m}}\rangle=0$ requires a quasistatic
insertion and removal of the system-demon coupling. Equilibrium would
be maintained throughout, implying, of course, that the system and
demon are uncorrelated at $t=0$; their joint pdf returns to $p_{{\rm eq}}^{x}(x_{0})p_{{\rm eq}}^{y}(y_{0})$,
where $x_{0}$ and $y_{0}$ denote the displacements at $t=0$. This
is not a useful measurement procedure: it is as though it never took
place. A useful measurement requires the input of positive work, on
average.

We therefore consider instead a nonquasistatic procedure such that
the system and device are described by a nonequilibrium pdf at $t=0$
represented by $p(x_{0},y_{0})$. The correlation between system and
demon can then be most usefully expressed in terms of the so-called
\emph{mutual information} \cite{Sagawa08,Sagawa09,Sagawa12,Sagawa12b}:
\begin{equation}
I_{m}=\int dy_{0}dx_{0}\,p(x_{0},y_{0})\ln\frac{p(x_{0},y_{0})}{p^{x}(x_{0})p^{y}(y_{0})},\label{eq:12}
\end{equation}
where $p^{x}(x_{0})=\int dy_{0}\,p(x_{0},y_{0})$ and $p^{y}(y_{0})=\int dx_{0}\,p(x_{0},y_{0})$
are the pdfs of system and demon at $t=0$. If $p(x_{0},y_{0})$ is
separable such that system and demon are statistically independent,
the mutual information vanishes; otherwise it is positive.

Since $\langle W_{{\rm m}}\rangle>0$ for a nonquasistatic process
where the Hamiltonian returns to its initial form, a positive work
is required, on average, to establish the nonequilibrium distribution
$p(x_{0},y_{0})$. We shall focus attention on what would appear to
be the least irreversible but still useful measurement procedure (indeed
since there is zero mean total entropy production it can be considered
to be reversible). We introduce the coupling quasistatically but remove
it instantaneously at $t=0$: $K(t)$ evolves extremely slowly from
zero at $t=-\tau_{{\rm m}}$ until $t=0$, (implying that $\tau_{{\rm m}}$
is very large) at which time it is equal to $\kappa>0$, and then
is abruptly taken to zero. The mean work performed during the quasistatic
process is the free energy change associated with the introduction
of the system-demon coupling \cite{fordbook}, and the mean work performed
in the abrupt decoupling is just the mean change in potential energy
of the system and demon at that instant. We therefore write the mean
work of measurement as $\langle W_{{\rm m}}^{{\rm qi}}\rangle=\Delta F_{{\rm m}}(\kappa)-\int dx_{0}dy_{0}\,p(x_{0},y_{0})H_{{\rm m}}(x_{0},y_{0},\kappa)$
where $\Delta F_{{\rm m}}(\kappa)$ is the free energy change associated
with the introduction of the coupling oscillator with spring constant
$\kappa$, with the label qi indicating that the work arises from
a measurement protocol of quasistatic coupling and instantaneous decoupling.
If the coupling stage were conducted nonquasistatically, the second
law tells us that the mean work would be greater than this, of course.

For this measurement procedure, the system and device are in equilibrium
just prior to $t=0$, and their joint pdf would be unchanged by the
decoupling, though it then becomes a nonequilibrium state. The pdf
after measurement is
\begin{eqnarray}
p(x_{0},y_{0}) & = & p_{{\rm eq}}^{x}(x_{0})p_{{\rm eq}}^{y}(y_{0})\nonumber \\
 & \times & \exp[(\Delta F_{{\rm m}}(\kappa)-H_{{\rm m}}(x_{0},y_{0},\kappa))/kT],\quad\label{eq:13}
\end{eqnarray}
so the mutual information characterising the measurement is
\begin{eqnarray}
 & I_{m} & =\int dy_{0}dx_{0}\,p(x_{0},y_{0})\biggl(\ln\frac{p_{{\rm eq}}^{x}(x_{0})p_{{\rm eq}}^{y}(y_{0})}{p^{x}(x_{0})p^{y}(y_{0})}\nonumber \\
 &  & \quad\qquad+[\Delta F_{{\rm m}}(\kappa)-H_{{\rm m}}(x_{0},y_{0},\kappa)]/kT\biggr)\nonumber \\
 &  & =\langle W_{{\rm m}}^{{\rm qi}}\rangle/kT-D_{{\rm KL}}(p^{y}\vert\vert p_{{\rm eq}}^{y})-D_{{\rm KL}}(p^{x}\vert\vert p_{{\rm eq}}^{x}),\label{eq:400}
\end{eqnarray}
where we introduce Kullback-Leibler divergences or \emph{relative
entropies} between the pdfs $p^{x}$ and $p_{{\rm eq}}^{x}$, and
$p^{y}$ and $p_{{\rm eq}}^{y}$, defined by
\begin{equation}
D_{{\rm KL}}(p^{x}\vert\vert p_{{\rm eq}}^{x})=\int dx_{0}\,p^{x}(x_{0})\ln\frac{p^{x}(x_{0})}{p_{{\rm eq}}^{x}(x_{0})},\label{eq:16}
\end{equation}
and similarly for the demon. A relative entropy quantifies the difference
between two pdfs; it may be shown that $D_{{\rm KL}}(p\vert\vert p^{\prime})\ge0$,
and that it vanishes when pdfs $p$ and $p^{\prime}$ are identical.

\begin{figure}
\begin{centering}
\includegraphics[width=0.79\columnwidth]{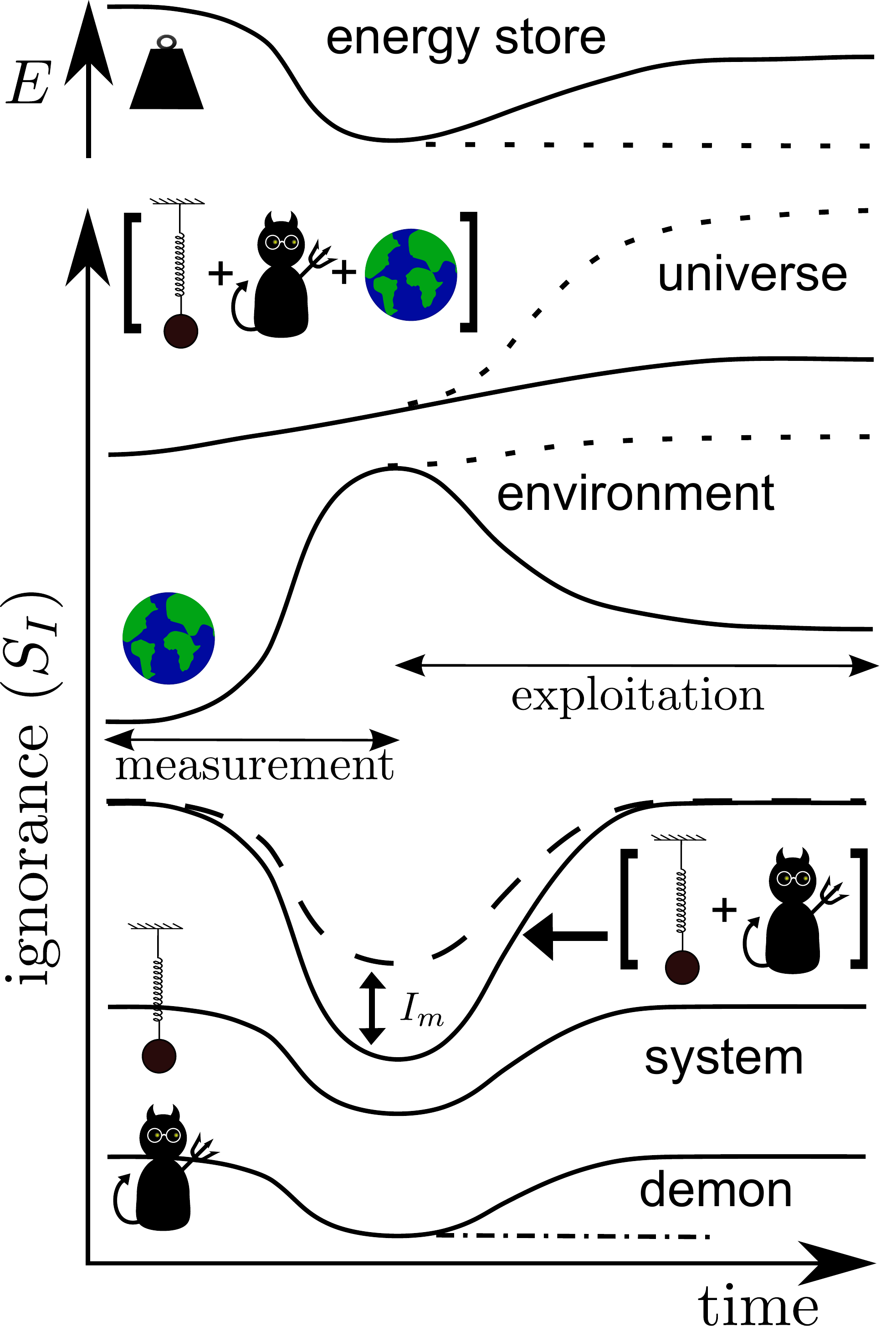}
\par\end{centering}

\caption{Ignorance management during measurement and exploitation. The coupling
between system and demon brings about a reduction in their Shannon
entropies $S_{I}$: these are shown separately, and their sum is given
as the long-dashed curve. The Shannon entropy of the system and demon
combined is less than this sum, the difference being the mutual information
$I_{m}$ that is a reflection of their correlation (see Eq. (\ref{eq:26})).
The uncertainty in the microstate of the environment increases during
the measurement phase, as a consequence of heat transfer, and the
net change in our ignorance of the microstate of the combined system/demon/environment
(or universe) is positive. A reduction in the mean energy $E$ held
in the potential energy store drives these changes. If no exploitation
operations are carried out, the potential energy store is not replenished
and the nonequilibrium states of the system and demon simply relax
after decoupling, the irreversibility of which is reflected in increased
uncertainty that ultimately accumulates in the environment (short-dashed
curves). However, if suitable exploitation operations are invoked,
heat can be drawn from the environment and returned to the potential
energy store and there is less overall thermodynamic entropy production
(solid curves during exploitation phase). We also note that the post-measurement
nonequilibrium state of the demon could be preserved (dash-dotted
line) for later treatment, whether it be exploitation or simply relaxation.
\label{fig:Ignorance-management-during}}
\end{figure}

Both the system and the demon are left in nonequilibrium statistical
states after the measurement procedure considered, so both the relative
entropies are nonzero. Intuitively, this means that the Shannon entropies
of both system and demon have been \emph{reduced} with respect to
their initial equilibrium states, as illustrated at the bottom of
Figure \ref{fig:Ignorance-management-during}, where we sketch the
process of ignorance management. The Figure illustrates that the Shannon
entropy of the combined system and demon is less than the sum of their
respective Shannon entropies, because of the correlation between them.
As in Figure \ref{fig:Sharpened-demon}, the Shannon entropy of the
environment rises as a consequence of the reduction in ignorance of
system and demon, and the Shannon entropy of the universe increases
as a consequence of mean total stochastic entropy production. These
events are driven by the depletion, on average, of the potential energy
store. We rely on physical intuition here, but an explicit example
to illustrate these assertions is given in Appendix \ref{sec:mathematical-detail}.

According to Eq. (\ref{eq:400}), the mean work of measurement is
therefore related to a set of statistical correlations: the mutual
information and two relative entropies. The technical terminology
can be distracting, but the essential meaning is clear: correlation
is created by (mean) work.

If no action were taken to exploit the measurement, the system and
demon, assuming they remained coupled to the environment, would then
simply relax back to their respective equilibrium states, accompanied
by the positive mean production of total stochastic entropy involving
heat exchange with the environment. Measurement does not \emph{have}
to be followed by further action, but the energy taken from the store
to perform the measurement would then have been wasted. However, a
demon can inform an exploitation or feedback scheme that might return
some of this energy, and we consider this next.

\subsection{Exploitation\label{sub:Exploitation}}

The development of our ignorance of various components of the world
during the exploitation of a measurement is sketched in Figure \ref{fig:Ignorance-management-during}.
The range of what is possible can be determined by some mathematical
analysis, the elements of which we describe shortly, but the key outcomes
are essentially those that follow from the measurement in Figure \ref{fig:demon-at-work}:
environmental heat is converted into stored potential energy, meaning
that environmental Shannon entropy can be reduced, while the Shannon
entropies of system and demon are returned to their initial values,
during all of which the universe remains subject to the second law.

We need to characterise the work extraction made possible by the demon's
measurement of the system, and the key to understanding this is to
recognise, following Sagawa and Ueda \cite{Sagawa12}, that systems
subject to feedback satisfy modified forms of the Jarzynski equality.

We shall consider a simple feedback procedure where the demon's displacement
$y_{0}$ at $t=0$ is the input that specifies a subsequent protocol
of manipulation of the system. The exploitation will take the form
of a time-dependent Hamiltonian $H_{{\rm e}}(x,t\vert y_{0})$ that
operates on the system, and we require it to start from zero at $t=0$
and vanish at $t=\tau_{{\rm e}}$ at the end of the exploitation period.

If we consider the system, for the moment, to be in equilibrium at
$t=0$, such that $p^{x}(x_{0})=p_{{\rm eq}}^{x}(x_{0})$, values
of the work $W_{{\rm e}}(y_{0})$ performed during realisations of
exploitation under this Hamiltonian will satisfy an unmodified Jarzynski
equality equivalent to Eq. (\ref{eq:6}):
\begin{eqnarray}
1 & = & \int dx_{0}\,p_{{\rm eq}}^{x}(x_{0})\langle\exp(-W_{{\rm e}}(y_{0})/kT)\rangle_{x_{0}},\label{eq:17}
\end{eqnarray}
where $\langle\exp(-W_{{\rm e}}(y_{0})/kT)\rangle_{x_{0}}$ is an
average over system trajectories that start at initial position $x_{0}$,
developing under dynamics that depend upon $y_{0}$. We now take an
average over $y_{0}$ with weighting $p^{y}(y_{0})$ to examine the
statistics of the work informed by a different exploitation Hamiltonian
for each measurement outcome. We write

\begin{equation}
\begin{aligned}1 & =\int dy_{0}\,p^{y}(y_{0})\,dx_{0}\,p_{{\rm eq}}^{x}(x_{0})\langle\exp(-W_{{\rm e}}(y_{0})/kT)\rangle_{x_{0}}\\
 & =\int dy_{0}dx_{0}\,p^{y}(y_{0})p^{x}(x_{0})\langle\exp(-W_{{\rm e}}(y_{0})/kT-\delta s_{{\rm sys}})\rangle_{x_{0}},
\end{aligned}
\label{eq:18}
\end{equation}
where $p^{x}(x_{0})$ is the actual distribution of $x_{0}$ at $t=0$,
typically differing from the equilibrium distribution $p_{{\rm eq}}^{x}(x_{0})$,
and we have introduced the associated difference in stochastic system
entropy $\delta s_{{\rm sys}}=-\ln p_{{\rm eq}}^{x}(x_{0})/p^{x}(x_{0})$.
We therefore obtain
\begin{eqnarray}
1 & = & \int dy_{0}dx_{0}\,p(x_{0},y_{0})\nonumber \\
 & \times & \langle\exp(-W_{{\rm e}}(y_{0})/kT-\ln\frac{p(x_{0},y_{0})}{p^{x}(x_{0})p^{y}(y_{0})}-\delta s_{{\rm sys}})\rangle_{x_{0}}\nonumber \\
 & = & \overline{\langle\exp(-W_{{\rm e}}(y_{0})/kT-I_{x_{0}y_{0}}-\delta s_{{\rm sys}})\rangle},\label{eq:19}
\end{eqnarray}
where the brackets in the final result denote an average over system
trajectories, with exploitation Hamiltonian conditioned on $y_{0}$,
that start at all possible $x_{0}$; and the bar indicates an average
over all values of the exploitation protocol label $y_{0}$, with
the statistics of these variables described by joint pdf $p(x_{0},y_{0})$,
and where $I_{x_{0}y_{0}}=\ln[p(x_{0},y_{0})/p^{x}(x_{0})p^{y}(y_{0})]$.

This has the consequence that
\begin{equation}
\int dy_{0}dx_{0}\,p(x_{0},y_{0})\left(\langle W_{{\rm e}}(y_{0})\rangle_{x_{0}}/kT+I_{x_{0}y_{0}}+\delta s_{{\rm sys}}\right)\ge0,\label{eq:20}
\end{equation}
which is simply a form of the second law written in terms of work,
mutual information and stochastic system entropy.

For clarification, we now write $p(x_{0},y_{0})=P(x_{0}\vert y_{0})p^{y}(y_{0})$,
where $P(x_{0}\vert y_{0})$ is a conditional probability density,
such that Eq. (\ref{eq:20}) becomes
\begin{align}
 & \overline{\langle W_{{\rm e}}(y_{0})\rangle}/kT+\int dy_{0}\,p^{y}(y_{0})\left(\int dx_{0}\,P(x_{0}\vert y_{0})I_{x_{0}y_{0}}\right)\nonumber \\
 & \qquad+\int dx_{0}\,p^{x}(x_{0})\ln p^{x}(x_{0})/p_{{\rm eq}}^{x}(x_{0})\ge0.\label{eq:21}
\end{align}
The second term involves the relative entropy between distributions
$P(x_{0}\vert y_{0})$ and $p^{x}(x_{0})$:
\begin{eqnarray}
D_{{\rm KL}}(P\vert\vert p^{x}) & = & \int dx_{0}\,P(x_{0}\vert y_{0})I_{x_{0}y_{0}}\nonumber \\
 & = & \int dx_{0}\,P(x_{0}\vert y_{0})\ln(P(x_{0}\vert y_{0})/p^{x}(x_{0})),\quad\quad\label{eq:22}
\end{eqnarray}
and the final term is the relative entropy between $p^{x}(x_{0})$
and $p_{{\rm eq}}^{x}(x_{0})$.

We note that the mutual information between system and demon, introduced
earlier in Eq. (\ref{eq:12}), is the average over $y_{0}$ of the
relative entropy $D_{{\rm KL}}(P\vert\vert p^{x})$:
\begin{align}
I_{m} & =\int dy_{0}\,p^{y}(y_{0})D_{{\rm KL}}(P\vert\vert p^{x})=\int dy_{0}dx_{0}\,p(x_{0},y_{0})I_{x_{0}y_{0}}\nonumber \\
 & =\int dy_{0}dx_{0}\,p(x_{0},y_{0})\ln\frac{p(x_{0},y_{0})}{p^{x}(x_{0})p^{y}(y_{0})},\label{eq:23}
\end{align}
so that we can write Eq. (\ref{eq:20}) as
\begin{equation}
\overline{\langle W_{{\rm e}}(y_{0})\rangle}/kT+I_{m}+D_{{\rm KL}}(p^{x}\vert\vert p_{{\rm eq}}^{x})\ge0.\label{eq:24}
\end{equation}

It has been noted \cite{Sagawa08,Sagawa09,Sagawa12,Sagawa12b} that
this result demonstrates that the mean exploitation work performed
on the system, averaged over system trajectories taken as well as
exploitation protocols identified by $y_{0}$, could be negative (corresponding
to a positive mean transfer to the potential energy store) since both
the mutual information $I_{m}$ and the relative entropy $D_{{\rm KL}}(p^{x}\vert\vert p_{{\rm eq}}^{x})$
are positive. Such an outcome would require a carefully designed set
of exploitation procedures, tailored to the outcome of the measurement
\cite{Sagawa12,Abreu11}.

Nevertheless, our objective is to combine Eqs. (\ref{eq:400}) and
(\ref{eq:24}) to notice that
\begin{equation}
\overline{\langle W_{{\rm e}}(y_{0})\rangle}+\langle W_{{\rm m}}^{{\rm qi}}\rangle\ge kT\,D_{{\rm KL}}(p^{y}\vert\vert p_{{\rm eq}}^{y})\ge0,\label{eq:25}
\end{equation}
so for a measurement protocol whereby a device is quasistatically
connected and then instantaneously decoupled, followed by a measurement-dependent
exploitation protocol, the mean extracted work $-\overline{\langle W_{{\rm e}}(y_{0})\rangle}$
is never greater than the mean work of measurement $\langle W_{{\rm m}}^{{\rm qi}}\rangle$.
The potential energy store never profits, on average, from the sequence
of events, as we illustrate through its evolution at the top of Figure
\ref{fig:Ignorance-management-during}. In the sense that it refers
to expected or mean behaviour, the nature of the second law in stochastic
thermodynamics is secure, at least for the specific measurement and
exploitation procedures we have considered. The law, and specifically
Eq. (\ref{eq:25}), is a statement about the unlikelihood of a successful
conversion of heat into work, even with feedback control.

We should note that this outcome is inevitable given that the mean
total stochastic entropy production is obliged to increase for any
nonquasistatic process modelled within a framework of stochastic thermodynamics,
whatever efforts are made by the demon in his channelling of feedback.
But we should also recognise that there are certainly realisations
of the process where the total stochastic entropy production is negative,
such that the store receives more energy during the exploitation of
a measurement than it had to provide in the making of the measurement.
It is just that these cases are lucky outcomes.

The demon could be frozen in its post-measurement microstate and replaced
by an equilibrated demon; or returned irreversibly to equilibrium
by thermalisation; or reused from its nonequilibrium state to perform
another measurement. However, it might also be exploited to replenish
the potential energy store. It may be shown that the mean work, in
units of $kT$, that may be returned to the store cannot be greater
than the relative entropy $D_{{\rm KL}}(p^{y}\vert\vert p_{{\rm eq}}^{y})$,
if the demon is treated appropriately, in which case the mean work
$\langle W_{{\rm e}}^{{\rm d}}\rangle$ done by the store on the demon
during such post-measurement processing is limited by $-\langle W_{{\rm e}}^{{\rm d}}\rangle\le kT\,D_{{\rm KL}}(p^{y}\vert\vert p_{{\rm eq}}^{y})$.
 In the light of this result, we could express Eq. (\ref{eq:25})
in the form
\begin{equation}
\overline{\langle W_{{\rm e}}(y_{0})\rangle}+\langle W_{{\rm e}}^{{\rm d}}\rangle+\langle W_{{\rm m}}^{{\rm qi}}\rangle\ge0,\label{eq:25b}
\end{equation}
which makes it absolutely clear that the average transfer of energy
at the end of the process is from the store to the environment.

\section{Rules for demons\label{sec:Rules-for-demons}}

At this point we reflect on the explicit and implicit rules for demons
that we seem to have employed in our considerations. If there were
no rules and no constraints, it would actually be quite easy to design
a scheme to guarantee the extraction of energy from a heat bath and
convert it into work. The emphasis here is on successful conversion
\emph{on} \emph{average}: it is of course quite expected in stochastic
thermodynamics that fluctuations will occur, whatever the rules.

For example, we could eliminate the demon in our example and allow
the system to exploit its own circumstances. Having equilibrated with
the heat bath, the system might autonomously and automatically invoke
a shift in its tethering point and transfer the potential energy difference
into the store. We would then wait for the system to re-equilibrate
with the heat bath until it is ready for the next extraction.

But, clearly, these dynamics do not generate a thermodynamic system
that has an equilibrium state when placed in contact with a heat bath.
It is a self-sorting process that we might consider makes an illegal
challenge to the second law. Pedantically, perhaps such dynamical
schemes are examples of a successful demon at work, but they do not
model what we would normally regard as thermodynamic systems. We are
thus guided towards setting strict criteria that define the nature
of the problem.

Another simple example of an autonomous self-sorting dynamical system
is a particle that leaves a trail of regions that it has passed through,
and from which it is barred from visiting again in the future. The
volume available to the particle is progressively diminished and the
uncertainty corresponding to its position in the space is reduced.
The system has sorted itself and reduced its own system entropy. Such
schemes are actually employed in the evaluation of potentials of mean
force in molecular dynamics studies where the approach is known as
metadynamics \cite{Laio02}, but again we do not regard this as an
example of a system that exhibits traditional thermodynamic behaviour.

There has to be an external intervention to initiate the exploitation
of a thermodynamic system, and that is precisely the role of the demon.
But the demon's dynamics cannot be self-sorting, if we assume that
in practical situations he is also a thermodynamic system. The demon
appears to be restricted to controlling the sorting of some other
part of the world, and is barred from exercising any control over
his own development. In order to make a measurement, a coupling between
the demon and system has to be made, and this must be removed by the
time the exploitation is initiated in order to avoid any possibility
of self-sorting. This is our principal rule for demons.

If the microstate of a demon can be used to inform the exploitation
of a system, might the system microstate be used to exploit the demon?
But it can perhaps be argued that a distinction should be made between
a system and a demon, such that a system simply cannot be used to
inform any subsequent action: that by definition this ability is possessed
only by the demon.

But we can then imagine two demons, each able to inform the exploitation
of the other. This might be just another case of self-sorting: a feasible
set of dynamics but somehow breaking the rules for demons. However,
let us imagine how this might proceed. We use the notation of our
example, but allow the system to inform an exploitation procedure
that applies to the demon. According to Eq. (\ref{eq:24}) we had
\begin{equation}
\overline{\langle W_{{\rm e}}(y_{0})\rangle}/kT+I_{m}+D_{{\rm KL}}(p^{x}\vert\vert p_{{\rm eq}}^{x})\ge0,\label{eq:35}
\end{equation}
that constrained the mean exploitation work applied to the system,
when controlled by the demon displacement $y_{0}$ at $t=0$. In our
two-demon situation there is a corresponding expression for the mean
exploitation work obtained by manipulation of the demon spring strength
and tethering point, conditioned on the microstate of the system:
\begin{equation}
\overline{\langle W_{{\rm e}}(x_{0})\rangle}/kT+I_{m}+D_{{\rm KL}}(p^{y}\vert\vert p_{{\rm eq}}^{y})\ge0,\label{eq:36}
\end{equation}
and using Eq. (\ref{eq:400}) we can then write
\begin{equation}
\overline{\langle W_{{\rm e}}(x_{0})\rangle}+\overline{\langle W_{{\rm e}}(y_{0})\rangle}+\langle W_{{\rm m}}^{{\rm qi}}\rangle+kT\ I_{m}\ge0,\label{eq:37}
\end{equation}
so that the total mean work performed on the system (the depletion
of the potential energy store) is greater than $-kTI_{m}$, and therefore
potentially negative.

So it is easy to construct challenges to the second law by imagining
dynamical systems that evolve in particular ways, and perhaps these
possibilities simply tell us how to specify the rules that should
be applied to demons to ensure that they are ultimately unsuccessful.
We might declare that any scheme that, on average, converts heat into
work through autonomous dynamics, is not working in the `spirit' of
the challenge to thermodynamics posed by Maxwell's demon: indeed that
the dynamics discount thermodynamic behaviour in the first place.
On the same evidence we might conclude, with Maxwell himself, that
there are feasible physical situations that perform sorting, suggesting
there are no inviolable laws against operations that would include
the conversion of heat into work, only practical difficulties.

The main point is that these thermodynamic issues are somewhat clarified
when a dynamical framework is employed. A further point is that the
second law that is to be challenged is not a rigid exclusion of behaviour,
but rather a statement of expectation. As we noted earlier, stochastic
dynamics provide a natural framework for the evolution of a system
coupled to an environment, so it is easy to accept that fluctuations
in thermodynamic outcomes are possible. No rules on demons can exclude
these possibilities.

\section{Conclusions\label{sec:Conclusions}}

In this article we have summarised a position that can be taken on
Maxwell's demon that arises from explicit modelling of the process
of measurement and exploitation within a framework of stochastic thermodynamics.

The demon has received repeated attention since his activities were
imagined by Maxwell nearly 150 years ago, and the precise meaning
of the second law that he challenges has occasionally shifted. But
perhaps not enough consideration has been given to the dynamics underpinning
his actions. In stochastic thermodynamics we have the advantage that
the dynamics of measurement, exploitation and equilibration are all
well specified, as is the exact meaning of entropy production and
the second law. On the other hand, certain dynamical assumptions are
introduced, such as the explicit breakage of time reversal symmetry
or simple forms of the dissipative and noise terms, and we appear
to have to define rules for admissible exploitation strategies. Nevertheless,
we can present these ingredients in a transparent fashion.

We have a picture where the demon monitors a measuring device (or
\emph{is }the device) that can be coupled dynamically to an evolving
system, and the microstate of the device may be used to control a
subsequent programme of feedback on the system dynamics. Both demon/device
and system are influenced by noise from the environment. The intention
is that we are modelling a physical system whose microscopic configuration
is inaccessible except through the microstate taken by a measuring
device. We regard it as inadmissible that the microstate of the system
might be used to control feedback that acts upon itself: dynamics
of this kind would be essentially `self-sorting' with a natural tendency
to evolve towards an attractor, in contrast to dynamics that display
a sensitivity to initial conditions.

The coupling between system and device is switched on and off, requiring
work to be done (taken from a potential energy store), but yielding
a statistical correlation represented by a mutual information between
the system and device. The exploitation process on the system, informed
by the demon, can then transfer heat from the environment into potential
energy in the store, while returning the system to its condition prior
to the measurement. The post-measurement device can also be exploited
to return some energy to the store but the procedure followed need
not be tailored to its exact microstate: it is not feedback. However,
the \emph{average} work done by the potential energy store on the
system and demon/device in performing the measurement and exploiting
the result \emph{cannot} be negative, as demonstrated in the analysis
considered in Section \ref{sec:A-simple-model}. The origin of this
statement is the second law of thermodynamics, in the form of an integral
fluctuation relation for stochastic entropy production, adapted to
the situation of measurement and feedback by Sagawa and Ueda \cite{Sagawa12}.

Stochastic dynamics and thermodynamics allow us to model individual
realisations of a process, and to demonstrate that there are fluctuations
in behaviour. The dynamical equations are stochastic, reminding us
that we deal with uncertainty in evolution and that unexpected outcomes
are possible, such as a violation of Kelvin's statement of the second
law, a transformation of energy from heat into work. However, such
cases are outweighed in probability by examples where the flow is
in the opposite direction, and law-breaking realisations are rarer
as the system becomes larger and more complex. We can accommodate
the possibility that Kelvin's statement might be violated by small
systems over short periods of time, while maintaining the usual restrictions
at macroscopic scales.

The second law in its traditional, rigid form can be broken by fluctuation:
the role of the demon is to attempt to break it on average. But we
impose rules on demons that might seem unfair, in that any successful
strategy can be declared to be brought about by illegal dynamics.
It is important to be clear that such rules exist. By requiring that
feedback on a system can only be channelled through a device or demon
that is coupled to the system and then decoupled, with mechanical
consequences, we can eliminate, or at least categorise, puzzling counter-examples
to the second law, such as those involving the insertion of partitions
into cavities, and their manipulation in the knowledge that a particle
lies to one side or the other (the Szilard engine \cite{Szilard29},
illustrated in Figure \ref{fig:demon-at-work}). On the other hand,
we could take the point of view that sustained breakages of the second
law would not be surprising if we were allowed an unrestricted choice
of dynamics. But the more usual fundamental position is that we consider
thermodynamic phenomena to be underpinned by system dynamics that
are sensitive to initial conditions, and that it is most appropriate
to represent the behaviour using stochastic equations of motion that
tend to increase the uncertainty in the microstate. This being so,
self-sorting behaviour is excluded and the demon must ultimately fail.

The position just outlined might be contrasted with two earlier points
of view designed to demonstrate that the demon cannot succeed. We
start with Option 1: the viewpoint associated with the Bennett exorcism
of the demon \cite{Bennett82} and illustrated in Figure \ref{fig:Confused-demon}.
The steps in a process are as follows:

\begin{itemize}[noitemsep,leftmargin=*]
\item Measure the system and reduce its entropy (the demon discovers the speed of a gas particle in Maxwell's original thought experiment).
\item Exploit the measurement and cement the reduction in entropy (the demon manipulates a trapdoor and sorts the gas).
\item The world owes a debt to the Bank of Negative Entropy (the IOU is the demon's `memory' of past measurement).
\item Resetting the measuring device generates entropy to clear the debt of entropy to the Bank (the demon's memory is wiped).
\end{itemize}

A state of reduced entropy is granted on the understanding that when
the measuring device is later reset to a standard state, compensating
entropy is generated. But a second law that can be violated for an
indefinite period of time might be viewed as no law at all. Can we
accept this?

Then there is Option 2: an improved viewpoint that does address dynamical
evolution but which is implicitly deterministic in nature. It too
is illustrated in Figure \ref{fig:Confused-demon} but the uncertainty
acquired by the demon is classed as entropy. The steps are:

\begin{itemize}[noitemsep,leftmargin=*]
\item Measure system with a (possibly conservative) exchange of entropy between system and device.
\item Exploit the reduced system entropy, perhaps converting environmental heat into work, returning the system to its initial state.
\item The measuring device persists in a higher entropy state.
\item The world proceeds free of debt.
\end{itemize}

Post-measurement, the device has received an increase in entropy.
There is no need for a reset to satisfy the requirements of the second
law: entropy has been simply transferred. An eventual return of the
device to its original equilibrium state would simply pass the additional
entropy back into the environment.

And then there is Option 3: the viewpoint based on stochastic dynamics
and thermodynamics. It is illustrated in Figure \ref{fig:Sharpened-demon}.
The steps are:

\begin{itemize}[noitemsep,leftmargin=*]
\item Measure system with the input of work.
\item System and measuring device become correlated, and separately disturbed from equilibrium,  corresponding to reduced entropy, or less uncertainty in their microstate compared with the pre-measurement situation. The microstate of the environment becomes more uncertain.
\item Exploit measurement with the conversion of environmental heat into work, principally by manipulating the system but possibly the device as well.
\item The mean work extracted is less than the mean work input: this is the second law in this context.
\end{itemize}

Our ignorance is managed in Option 3, figuratively by the demon, in
a way that seems very different compared with Options 1 and 2. Measurement
leaves both system and device in less uncertain microstates, but the
environment is made more uncertain since, on average, there has to
be an increase in total stochastic entropy.

Kelvin's statement of the second law, when regarded as a restriction
on average behaviour, is safe in this framework since making the measurement
requires the \emph{upfront} performance of work. This is arguably
a much more acceptable framing of the law: it is temporally resilient
whatever the actions of the demon. In contrast to Option 1, we can
never get ahead in work terms; we cannot set aside the requirements
of the law for an indefinite period. A cycle does not have to be completed
and no loans have to be sought from the Bank of Negative Entropy.

In contrast with Option 2, we have in Option 3 a realistic dynamical
framework representing the act of measurement that includes the stochasticity
of environmental interaction. According to this picture, there is
nothing particularly special about the demon: he makes enquiries,
and performs actions depending on the response. In short, he merely
behaves like a tiny version of one of us.

\subsection*{Acknowledgements}

This work was partially supported by the COST Action MP1209 and the
EPSRC Network Plus on Emergence and Physics far from Equilibrium,
and I am grateful to Surani Gunasekera for assistance in its early
stages.

\appendix

\section{Example of evolution of ignorance during and after measurement\label{sec:mathematical-detail}}

We provide mathematical support for the claims made in Section \ref{sub:Measurement}
with regard to the evolution of our ignorance of the system, demon
and environment microstates.

The correlation or mutual information established after an investment
of energy in the coupling between system and demon suggested that
our ignorance about the microstate of the system-demon composite had
been reduced,  which is intuitively reasonable since it ties in with
the idea that a measurement has been made.  We can make the measurement
process more explicit using the analytical tractability of the system
and demon.

We can express the mutual information in terms of Shannon entropies:
\begin{eqnarray}
I_{m} & = & \int dy_{0}dx_{0}\,p(x_{0},y_{0})\ln\frac{p(x_{0},y_{0})}{p^{x}(x_{0})p^{y}(y_{0})}\nonumber \\
 & = & -S_{I}^{{\rm s+d}}+S_{I}^{{\rm s}}+S_{I}^{{\rm d}},\label{eq:26}
\end{eqnarray}
where the uncertainty in the microstate of the system-demon composite
is
\begin{equation}
S_{I}^{{\rm s+d}}=-\int dy_{0}dx_{0}\,p(x_{0},y_{0})\ln p(x_{0},y_{0}).\label{eq:27}
\end{equation}
The evolution of this quantity was sketched in Figure \ref{fig:Ignorance-management-during}.
Assuming the system and demon are initially in their respective equilibrium
states, the change in Shannon entropy of the system-demon composite
after the measurement is
\begin{equation}
\Delta S_{I}^{{\rm s+d}}=S_{I}^{{\rm s+d}}-S_{I,{\rm eq}}^{{\rm s}}-S_{I,{\rm eq}}^{{\rm d}}=-I_{m}+\Delta S_{I}^{{\rm s}}+\Delta S_{I}^{{\rm d}},\label{eq:28}
\end{equation}
having introduced the change in system Shannon entropy $\Delta S_{I}^{{\rm s}}=S_{I}^{{\rm s}}-S_{I,{\rm eq}}^{{\rm s}}=-\int dx_{0}\,p^{x}(x_{0})\ln p^{x}(x_{0})+\int dx\,p_{{\rm eq}}^{x}(x)\ln p_{{\rm eq}}^{x}(x)$,
and a similar expression for $\Delta S_{I}^{{\rm d}}$, where the
integration variable in the second term refers to a situation at $t=-\tau_{{\rm m}}$.

In order to understand the meaning of Eq. (\ref{eq:28}) we need to
determine the signs of the terms on the right hand side. For the uncoupled
demon we have $K_{y}=\kappa_{y}$ and an equilibrium distribution
$p_{{\rm eq}}^{y}(y)=(\kappa_{y}/2\pi kT)^{1/2}\exp(-\kappa_{y}y^{2}/2kT)$.
The harmonic interactions and assumed equilibrium at $t=-\tau_{{\rm m}}$
imply that the pdf of device and system takes a gaussian form throughout
the measurement interval, and specifically at $t=0$. After decoupling,
with system spring strength $K_{x}=\kappa_{x}$, we expect to find
that
\begin{eqnarray}
p^{y}(y_{0}) & = & \int dx_{0}\,p(x_{0},y_{0})\nonumber \\
 & \propto & \int dx_{0}\exp\left[-\frac{\kappa_{y}y_{0}^{2}}{2kT}-\frac{\kappa_{x}x_{0}^{2}}{2kT}-\frac{\tilde{K}_{0}(y_{0}-x_{0})^{2}}{2kT}\right]\nonumber \\
 & = & (\kappa_{{\rm eff}}^{y}/2\pi kT)^{1/2}\exp(-\kappa_{{\rm eff}}^{y}y_{0}^{2}/2kT),\label{eq:30}
\end{eqnarray}
where $\tilde{K}_{0}$ is a parameter to be determined, but required
to be equal to $K(0)=\kappa$ if the system and demon are in thermal
equilibrium prior to decoupling; and where $\kappa_{{\rm eff}}^{y}=(\kappa_{x}\kappa_{y}+\tilde{K}_{0}(\kappa_{x}+\kappa_{y}))/(\tilde{K}_{0}+\kappa_{x})$.
We then have
\begin{eqnarray}
D_{{\rm KL}}(p^{y}\vert\vert p_{{\rm eq}}^{y}) & = & \!\int dy_{0}\,p^{y}(y_{0})\ln\frac{p^{y}(y_{0})}{p_{{\rm eq}}^{y}(y_{0})}\nonumber \\
 & = & \!\int dy_{0}\,p^{y}(y_{0})\!\left[\frac{\kappa_{y}y_{0}^{2}}{2kT}-\frac{\kappa_{{\rm eff}}^{y}y_{0}^{2}}{2kT}-\frac{1}{2}\ln\!\left(\frac{\kappa_{y}}{\kappa_{{\rm eff}}^{y}}\right)\right]\nonumber \\
 & = & \frac{1}{2}\left[\left(\frac{\kappa_{y}}{\kappa_{{\rm eff}}^{y}}-1\right)-\ln\left(\frac{\kappa_{y}}{\kappa_{{\rm eff}}^{y}}\right)\right]\ge0,\label{eq:31}
\end{eqnarray}
and
\begin{eqnarray}
\frac{\kappa_{{\rm eff}}^{y}}{\kappa_{y}} & = & 1+\frac{\tilde{K}_{0}\kappa_{x}}{(\tilde{K}_{0}+\kappa_{x})\kappa_{y}}\ge1.\label{eq:32}
\end{eqnarray}
Since $\kappa_{{\rm eff}}^{y}\ge\kappa_{y}$ the displacement of
the demon is more narrowly distributed after the measurement than
before. This is consistent with
\begin{eqnarray}
\Delta S_{I}^{{\rm d}} & = & S_{I}^{{\rm d}}-S_{I,{\rm eq}}^{{\rm d}}\nonumber \\
 & = & -\int dy_{0}\,p^{y}(y_{0})\ln p^{y}(y_{0})+\int dy\,p_{{\rm eq}}^{y}(y)\ln p_{{\rm eq}}^{y}(y)\nonumber \\
 & = & -\int dy_{0}\,p^{y}(y_{0})\left[-\frac{\kappa_{{\rm eff}}^{y}y_{0}^{2}}{2kT}+\frac{1}{2}\ln\left(\frac{\kappa_{{\rm eff}}^{y}}{2\pi kT}\right)\right]\nonumber \\
 &  & +\int dy\,p_{{\rm eq}}^{y}(y)\left[-\frac{\kappa_{y}y^{2}}{2kT}+\frac{1}{2}\ln\left(\frac{\kappa_{y}}{2\pi kT}\right)\right]\nonumber \\
 & = & \frac{1}{2}\ln\left(\frac{\kappa_{y}}{\kappa_{{\rm eff}}^{y}}\right)\le0,\label{eq:33}
\end{eqnarray}
meaning that the uncertainty attached to the demon decreases upon
making the measurement. By similar considerations the same can be
said for the system, specifically $p^{x}(x_{0})\propto\exp(-\kappa_{{\rm eff}}^{x}x_{0}^{2}/2kT)$
with
\begin{equation}
\frac{\kappa_{{\rm eff}}^{x}}{\kappa_{x}}=1+\frac{\tilde{K}_{0}\kappa_{y}}{(\tilde{K}_{0}+\kappa_{y})\kappa_{x}}\ge1.\label{eq:33a-1}
\end{equation}
Upon further decoupled evolution in contact with the environment,
both system and demon would relax back to equilibrium and their Shannon
entropies would rise once again.

It is likely that the spring strength for the demon is smaller than
that of the system: the demon is imagined to adapt to the system microstate
and not the other way round. If $\kappa_{y}\ll\kappa_{x}$, then $\kappa_{{\rm eff}}^{y}\gg\kappa_{y}$
and $\kappa_{{\rm eff}}^{x}\approx\kappa_{x}$ such that $\vert\Delta S_{I}^{{\rm d}}\vert\gg\vert\Delta S_{I}^{{\rm s}}\vert$:
the measurement largely achieves a reduction of our ignorance of the
demon's microstate rather than of the system microstate: quite properly
since it is the increased clarity in the microstate of the demon that
informs the exploitation of the system.

Since both $\Delta S_{I}^{{\rm d}}$ and $\Delta S_{I}^{{\rm s}}$
are negative, we can elaborate on Eq. (\ref{eq:28}) and write
\begin{eqnarray}
\Delta S_{I}^{{\rm s+d}} & = & -I_{m}+\Delta S_{I}^{{\rm s}}+\Delta S_{I}^{{\rm d}}\le0,\label{eq:34}
\end{eqnarray}
 and since $\Delta S_{{\rm tot}}=\Delta S_{I}^{{\rm s+d}}+\Delta S_{{\rm env}}\ge0$,
the measurement process clearly produces a positive change in the
uncertainty of the environment, $\Delta S_{{\rm env}}\ge0$, suggesting
that heat on average passes from the potential energy store into the
environment: a dissipation. This is also illustrated in Figure \ref{fig:Ignorance-management-during}.
The more nonquasistatic the procedure, the greater the mean total
entropy production and heat transfer, and presumably the greater the
expected depletion of the energy store.

\begin{figure}
\begin{centering}
\includegraphics[width=1\columnwidth]{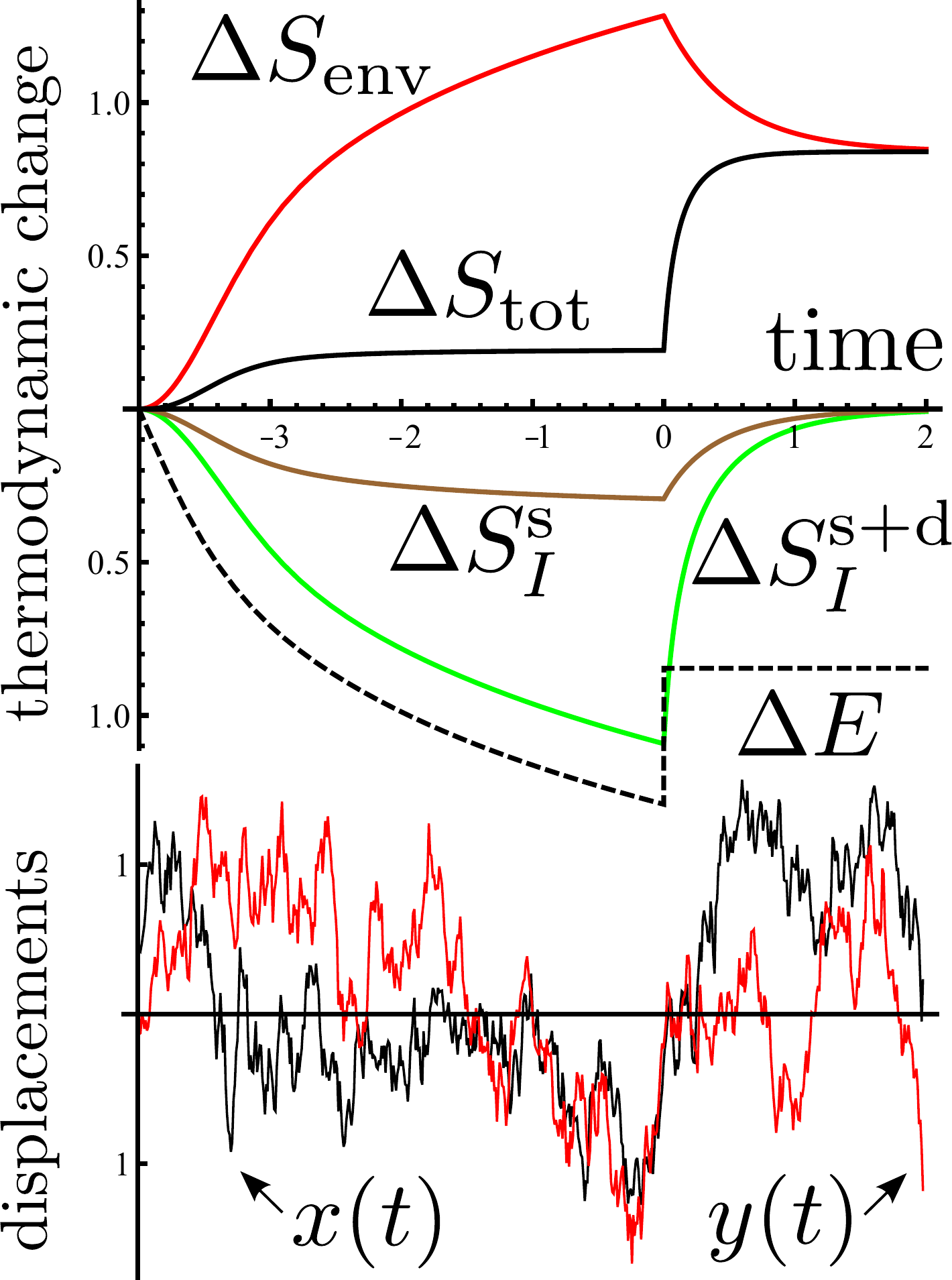}
\par\end{centering}

\caption{Evolution of thermodynamic quantities for an example of system-demon
coupling and decoupling without exploitation of the measurement. An
illustration of the stochastic dynamics of the system and demon is
shown at the bottom and details are given in the text. The behaviour
is consistent with the generic evolution of the same quantities sketched
in Figure \ref{fig:Ignorance-management-during}. \label{fig:evolution}}
\end{figure}

The story is illustrated explicitly in Figure \ref{fig:evolution}
using the dynamics of Eqs. (\ref{eq:8}) and (\ref{eq:9}) with $K_{x}=K_{y}=m=m^{\prime}=\gamma=\gamma^{\prime}=k=T=1$
and $K(t)=t+\tau_{{\rm m}}$ for $-\tau_{{\rm m}}\le t\le0$ with
$\tau_{{\rm m}}=4$ and $K(t)=0$ otherwise. A realisation of the
evolving displacements $x$ and $y$ of system and demon, respectively,
is also shown, to illustrate that the coupling term brings about a
correlation in their motion, that is lost for $t>0$ after the coupling
is removed.

The solution to the Fokker-Planck equation corresponding to the dynamics
takes the form of
\begin{eqnarray}
\!\!\!\!\!\!p(x,y,t) & = & \frac{1}{2\pi}[1+2\tilde{K}(t)]^{1/2}\nonumber \\
 & \times & \exp\left(-\frac{x^{2}}{2}-\frac{y^{2}}{2}-\frac{\tilde{K}(t)(y-x)^{2}}{2}\right),\label{eq:33a}
\end{eqnarray}
(recalling that certain parameters have been set to unity) where $\tilde{K}(t)$
is determined by
\begin{equation}
\frac{d\tilde{K}}{dt}=-2(\tilde{K}-K)(1+2\tilde{K}),\label{eq:33b}
\end{equation}
with $\tilde{K}(-\tau_{{\rm m}})=0$, so that in the quasistatic limit
the pdf parameter $\tilde{K}(t)$ will mirror the evolution of the
coupling strength $K$. The parameter $\tilde{K}_{0}$ in Eq. (\ref{eq:30})
corresponds to $\tilde{K}(0)$. For the case considered, $\tilde{K}(t)$
rises quasilinearly to reach a value approaching four at $t=0$, and
then decays to zero over a time interval of order unity after the
abrupt decoupling.

The Shannon entropy of the demon evolves according to Eq. (\ref{eq:33})
with $\kappa_{y}=1$ and $\kappa_{{\rm eff}}^{y}=(1+2\tilde{K})/(1+\tilde{K})$,
and the Shannon entropy of the system is given by the same expression.
Both are illustrated by the curve labelled $\Delta S_{I}^{{\rm s}}$
in Figure \ref{fig:evolution}. The analysis allows us to express
the change in Shannon entropy of the composite of system and demon
as
\begin{equation}
\Delta S_{I}^{{\rm s+d}}=-\frac{1}{2}\ln(1+2\tilde{K}),\label{eq:33c}
\end{equation}
as can be seen in Figure \ref{fig:evolution}, and the mutual information
is given by
\begin{equation}
I_{m}=\frac{1}{2}\ln\left(1+\frac{\tilde{K}^{2}}{1+2\tilde{K}}\right).\label{eq:33d}
\end{equation}

The total stochastic entropy production for the dynamics of Eqs. (\ref{eq:8})
and (\ref{eq:9}) can be derived using methods given in \cite{SpinneyFord12b}.
Its increment is given by the It$\bar{{\rm o}}$-rules stochastic
differential equation \cite{Gardiner09}
\begin{eqnarray}
d\Delta s_{{\rm tot}} & = & 4(\tilde{K}-K)dt-(\tilde{K}-K)(1+2\tilde{K})(y-x)^{2}dt\nonumber \\
 &  & +(\tilde{K}-K)(x-y)(dx-dy),\label{eq:33e}
\end{eqnarray}
and it may be shown that the average rate of stochastic entropy production
is
\begin{eqnarray}
\frac{d\langle\Delta s_{{\rm tot}}\rangle}{dt} & = & \frac{4(\tilde{K}-K)^{2}}{1+2\tilde{K}},\label{eq:33f}
\end{eqnarray}
which, as required, is never negative.

The evolution of $\Delta S_{{\rm tot}}=\langle\Delta s_{{\rm tot}}\rangle$
is shown in Figure \ref{fig:evolution}, together with the change
in mean entropy of the environment defined by $\Delta S_{{\rm env}}=\Delta S_{{\rm tot}}-\Delta S_{I}^{{\rm s+d}}$.
The increase in $\Delta S_{{\rm env}}$ is associated with mean heat
flow to the environment through the dissipation of potential energy,
and this is illustrated by the change in the mean energy $\Delta E$
of the store: there is a decrease during system-demon coupling followed
by a smaller increase upon decoupling, and thereafter no replenishment.

The evolving quantities shown in Figure \ref{fig:evolution} are specific
examples of the more general behaviour sketched in Figure \ref{fig:Ignorance-management-during}.
Since we have not invoked an exploitation protocol to follow measurement
in our example, there is no recovery of potential energy to the store
and $\Delta S_{{\rm tot}}$ experiences a burst of production as the
system and demon relax to equilibrium. This is an example of the evolution
in Figure \ref{fig:Ignorance-management-during} indicated by the
short-dashed curves. Employing exploitation protocols such as those
suggested by Abreu and Seifert \cite{Abreu11}, for example, would
make use of the correlation and allow $\Delta S_{{\rm tot}}$, $\Delta S_{{\rm env}}$
and $\Delta E$ to follow behaviour more like the solid curves in
Figure \ref{fig:Ignorance-management-during}, but we shall not pursue
this.


%

\end{document}